\documentclass[floats,floatfix,showpacs,amssymb,prd,twocolumn,superscriptaddress,nofootinbib]{revtex4-1}

\usepackage{graphicx,epsf, epsfig, amssymb}% Include figure files
\usepackage{bm}
\usepackage{longtable}
\usepackage{color}
\usepackage[breaklinks]{hyperref}
\usepackage[caption=false]{subfig}
\usepackage{amsfonts,amsmath,amssymb,mathrsfs}

\def\nn{\nonumber}
\def\be{\begin{equation}}
\def\ee{\end{equation}}
\def\beq{\begin{eqnarray}}
\def\eeq{\end{eqnarray}}

\def\pa{\partial}
\def\ph{\varphi}
\def\th{\vartheta}

\def\cQ{{\cal Q}}

\begin{document}

\title{\large Scalar, Electromagnetic and Gravitational Perturbations
  \\of Kerr-Newman Black Holes in the Slow-Rotation Limit}

\author{Paolo Pani}\email{paolo.pani@ist.utl.pt}
\affiliation{CENTRA, Departamento de F\'{\i}sica, Instituto Superior
  T\'ecnico, Universidade T\'ecnica de Lisboa - UTL, Avenida~Rovisco Pais
  1, 1049 Lisboa, Portugal.}
\affiliation{Institute for Theory \& Computation, Harvard-Smithsonian
  CfA, 60 Garden Street, Cambridge, MA, USA}

\author{Emanuele Berti}\email{berti@phy.olemiss.edu}
\affiliation{Department of Physics and Astronomy, The University of
  Mississippi, University, MS 38677, USA.}
\affiliation{California Institute of Technology, Pasadena, CA 91109, USA}

\author{Leonardo Gualtieri}\email{Leonardo.Gualtieri@roma1.infn.it}
\affiliation{Dipartimento di Fisica, Universit\`a di Roma ``La
  Sapienza'' \& Sezione INFN Roma1, P.A. Moro 5, 00185, Roma, Italy.}

\date{\today} 

\begin{abstract}
  In Einstein-Maxwell theory, according to classic uniqueness
  theorems, the most general stationary black-hole solution is the
  axisymmetric Kerr-Newman metric, which is defined by three
  parameters: mass, spin and electric charge. The radial and angular
  dependence of gravitational and electromagnetic perturbations in the
  Kerr-Newman geometry do not seem to be separable. In this paper we
  circumvent this problem by studying scalar, electromagnetic and
  gravitational perturbations of Kerr-Newman black holes in the
  slow-rotation limit. We extend (and provide details of) the analysis
  presented in a recent {\em Letter} \cite{Pani:2013ija}. Working at
  linear order in the spin, we present the first detailed derivation
  of the axial and polar perturbation equations in the
  gravito-electromagnetic case, and we compute the corresponding
  quasinormal modes for any value of the electric charge. Our study is
  the first self-consistent stability analysis of the Kerr-Newman
  metric, and in principle it can be extended to any order in the
  small rotation parameter. We find numerical evidence that the axial
  and polar sectors are isospectral at first order in the spin, and
  speculate on the possible implications of this result.
\end{abstract}


\maketitle

\date{today}

%%%%%%%%%%%%%%%%%%%%%%%%%%%%%
\section{Introduction}
%%%%%%%%%%%%%%%%%%%%%%%%%%%%%

\subsection{Motivation and Background}

Classic uniqueness theorems (reviewed in, e.g.,
\cite{Chrusciel:2012jk}) guarantee that the Kerr-Newman (KN) metric
\cite{Newman:1965my} describes the most general stationary
electrovacuum solution in Einstein-Maxwell theory. The KN line element
\begin{eqnarray}
&&ds^2=-dt^2
+\Sigma\left(\frac{dr^2}{\Delta}+d\vartheta^2\right)
+(r^2+a^2)\sin^2\vartheta d\varphi^2\nonumber\\
&&+\frac{2Mr-Q^2}{\Sigma}
\left(dt-a\sin^2\vartheta d\varphi\right)^2\,,\label{KNmetric}
\end{eqnarray}
(where $\Sigma=r^2+a^2\cos^2\vartheta$, $\Delta=r^2+a^2-2Mr+Q^2$)
is characterized by 3 parameters: mass $M$, angular momentum $J=Ma$ and
electromagnetic charge $Q$. In the neutral case $(Q=0)$ the KN
solution reduces to the Kerr metric, whereas in the nonspinning limit
($J=0$) it reduces to the Reissner-Nordstr\"om (RN) metric.  When both
$Q\neq 0$ and $J\neq 0$ the spacetime is endowed with a magnetic
dipole moment, and it has the same gyromagnetic ratio $g=2$ as the
electron~\cite{Carter:1968rr}. This has led to some speculation that
the KN metric could be used as a classical model for elementary
particles \cite{Pekeris:1989mu}.

Most work in BH physics has focused on the Kerr metric, because
astrophysical BHs are likely to be electrically neutral. The reason is
that a BH of mass $M$ and charge $Q$ will not gravitationally accrete
particles of mass $m$ and charge $e$ as long as $eQ>Mm$ (in the
natural units $G=c=1$ used in this paper). Since $m/e\sim 10^{-21}$
for electrons, large BHs will hardly acquire any charge (see
e.g.~\cite{Gibbons:1975kk}).  Furthermore, in astrophysical
environments the electric charge is expected to be shorted out by the
surrounding plasma~\cite{Blandford:1977ds}.

Notwithstanding the fact that charge is unlikely to play a significant
role in astrophysics, the KN metric is a precious theoretical
laboratory to investigate the dynamics of Einstein-Maxwell theory in
curved spacetime, and the linearized dynamics of test fields on a KN
background have been intensively studied in the past.

The KN metric is the only nontrivial, asymptotically flat solution of
the Einstein-Maxwell system for which the geodesic and Klein-Gordon
equations can be solved by separation of
variables~\cite{Dadhich:2001sz}. The neutrino~\cite{Unruh:1973},
massive spin-$1/2$ \cite{Chandrasekhar:1976ap,Page:1976jj} and
Rarita-Schwinger \cite{TorresdelCastillo:1990aw}
equations in the KN metric are also known to be separable. The
separability of fermionic fields is related to the existence of a
generalized total angular momentum operator for the Dirac equation in
curved spacetime, that satisfies appropriate conservation
laws~\cite{Carter:1979fe,Carter:2006hj}. Scalar and Dirac
perturbations of a KN BH can therefore be treated using the same
general methods that apply to Kerr BHs. In particular, it is
straightforward to compute the quasinormal modes (QNMs) of KN BHs for
these classes of perturbations (see
\cite{Kokkotas:1999bd,Nollert:1999ji,Ferrari:2007dd,Berti:2009kk,Konoplya:2011qq} for
reviews on QNMs). For the KN metric, scalar QNMs were computed
in~\cite{Berti:2005eb} (see also Ref.~\cite{Konoplya:2013rxa} for a recent 
generalization to massive and charged scalar QNMs) and Dirac QNMs were computed
in~\cite{Jing:2005pk}.
The superradiant instability of massive scalar fields was studied
in~\cite{Furuhashi:2004jk}. Hartman {\it et al.} studied the
scattering of charged scalars and fermions in near-extremal KN
spacetimes~\cite{Hartman:2009nz}, complementing earlier work on the
so-called Kerr-Newman/Conformal Field Theory (KN/CFT)
conjecture~\cite{Hartman:2008pb}. Because the equations are separable,
(the absence of) superradiant effects~\cite{Teukolsky:1974yv} and
greybody factors for charged and massive Dirac particles in KN have
also been studied extensively
\cite{Wagh:1986cz,Pekeris:1989mu,Belgiorno:1998gj,Finster:1999ry,Finster:2000jz,
  Batic:2004sz,Batic:2005va,Batic:2006vs,Winklmeier:2006me,He:2006jv,Zhou:2008zzf}:
see \cite{Dolan:2009kj} for a good overview of work in this field, and
\cite{Belgiorno:2008hk} for a study of Dirac perturbations of KN BHs
in anti-de Sitter space.

Much less is known about the gravitational-electromagnetic sector of
KN perturbations. The reason is that most methods to compute QNMs,
greybody factors and scattering amplitudes (including the continued
fraction method~\cite{Leaver:1985ax} and the asymptotic iteration
method~\cite{Cho:2011sf}) rely on the separability of the perturbation
equations. Despite several
attempts~\cite{Dudley:1977zz,Dudley:1978vd,Bellezza:1984}, no one has
succeeded at separating the angular and radial dependence of the
gravitational-electromagnetic eigenfunctions.
Chandrasekhar's
monograph~\cite{Chandra} gives a comprehensive overview of this
long-standing unsolved problem.

A relatively small number of papers tried to address the problem of
the oscillations and stability of the KN metric under the combined
effect of electromagnetic and gravitational perturbations.
Mashhoon~\cite{Mashhoon:1985} used the analogy between geodesic
stability and QNMs first proposed by Goebel~\cite{1972ApJ...172L..95G}
(see also \cite{Cardoso:2008bp}) to estimate the QNMs in the eikonal
approximation and to argue that the KN metric should be stable. 
Dudley and Finley (\cite{Dudley:1977zz,Dudley:1978vd}, henceforth DF)
made a remarkable study of the separability of linear perturbations of
the solutions of the Einstein-Maxwell equations found by Pleb\'anski
and Demi\'anski \cite{Plebanski:1976gy}, which include all vacuum Type
D solutions in the Petrov classification. Due to nonseparability of
the perturbation equations, they could not treat
gravito-electromagnetic perturbations in a fully consistent approach.
In their work, DF ``either keep the geometry fixed and perturb the
electric field or, of more interest, keep the electric field fixed and
perturb the geometry''. This approach should be appropriate for values
of the charge $Q$ at most as large as the perturbations of the
spacetime metric.
Kokkotas first used the DF equation to compute the fundamental
gravitational QNM using WKB methods~\cite{Kokkotas:1993}. Later on,
Berti and Kokkotas \cite{Berti:2005eb} confirmed that the WKB
approximation is reasonably accurate for all values of the
dimensionless spin parameter ($\tilde{a}\equiv J/M^2$) and of the
charge $Q$ by comparing WKB results to a continued-fraction solution.
The main problem of the DF approach is not computational, but
physical. The DF equation does not treat the gravito-electromagnetic
coupling in a self-consistent way (for example, it does not reduce to
the well-known RN perturbation equations as $J\to 0$). Therefore it is
unclear whether it provides a correct description of gravitational and
electromagnetic perturbations of KN BHs.

In a recent {\em Letter}~\cite{Pani:2013ija}, we have presented the
first self-consistent study of the gravito-electromagnetic
perturbations of KN BHs. This paper complements and extends the
results of \cite{Pani:2013ija}, providing details of the derivation of
the perturbation equations and a more comprehensive set of numerical
results.

Our approach relies on a clear physical approximation, i.e. a
slow-rotation expansion of the perturbations of spinning
BHs~\cite{Pani:2012vp,Pani:2012bp}. The formalism to address this
problem was originally proposed in the context of slowly rotating
compact stars
\cite{ChandraFerrari91,Kojima:1992ie,1993ApJ...414..247K,Ferrari:2007rc},
and it can be extended (at least in principle) to any perturbative
order in the small rotation parameter. Within the slow-rotation
expansion (which is valid for any value of the BH charge $Q$) it is
possible to estimate truncation errors, e.g. by extending the
computation to the next order in rotation or by comparison with cases
where a nonperturbative solution is available (such as the case of
scalar perturbations considered below).

%\label{summary}
%%%%%%
%
\begin{figure*}[thb]
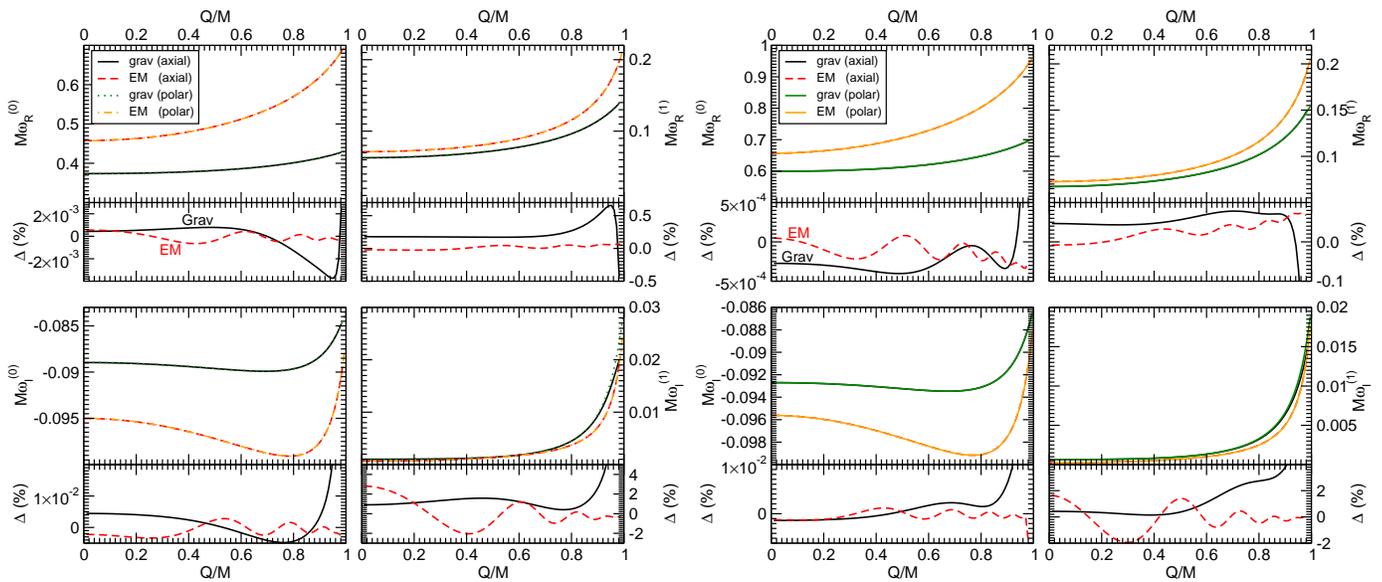

\begin{center}
\begin{tabular}{cc}
\epsfig{file=modes_l2.eps,width=9cm,angle=0,clip=true}&
\epsfig{file=modes_l3.eps,width=9cm,angle=0,clip=true}
\end{tabular}
\caption{Left (reproduced from Ref.~\cite{Pani:2013ija}): Zeroth-order
  (small left panels) and first-order (small right panels) terms of
  the slow-rotation expansion of the KN QNM frequencies
  [cf.~Eqs.~\eqref{wR} and \eqref{wI}]. All quantities are plotted as
  a function of $Q/M$, and they refer to the fundamental mode ($n=0$)
  with $\ell=2$.  The lower part of each panel shows the percentage
  difference between axial and polar quantities: our results are
  consistent with isospectrality to ${\cal O}(0.1\%)$ for the real
  part and to ${\cal O}(1\%)$ for the imaginary part of these
  modes. Right: same, but for fundamental modes with $\ell=3$.
\label{fig:modes}}
\end{center}
\end{figure*}
%%%%
%%%%%%%

\subsection{Executive Summary}

In the remainder of this Introduction we provide a short executive
summary of our main results, that is also meant as a guide to the
structure of the paper.

In Section~\ref{sec:formalism} we review our approach to separate the
scalar and the gravito-electromagnetic perturbations of a KN BH in the
slow-rotation limit~\cite{Kojima:1992ie,Pani:2012vp,Pani:2012bp}.
We derive the equations describing gravito-electromagnetic
oscillations in the slow-rotation approximation by linearizing the
Einstein-Maxwell equations with respect to both the amplitude of the
oscillation and the BH spin parameter $\tilde{a}\equiv J/M^2$. By
expanding the perturbations of the spacetime metric and of the
electromagnetic field in tensor spherical harmonics, we obtain a
coupled system of differential equations. 
Our main analytical result is the derivation of two sets of coupled,
second-order equations (one for the axial and one for the polar
sector, respectively) which fully describe gravito-electromagnetic
perturbations of a KN BH to first order in the spin:
%%%
\begin{eqnarray}
\hat {\cal D} Z_i^{\pm}&&\equiv V_0^{(i,\pm)}Z_i^{\pm}+m \tilde a\left[V_1^{(i,\pm)}Z_i^{\pm}+V_2^{(i,\pm)}{Z_i^{\pm}}'\right]\nn\\
&&+m \tilde a Q^2\left[W_1^{(i,\pm)} Z_j^{\pm}+W_2^{(i,\pm)}{Z_j^{\pm}}'\right],\label{eF}
\end{eqnarray}
%%%
Here $i,j=1,2$, $i\neq j$ and there is no sum over the indices $i,j$.
A prime denotes a derivative with respect to $r$, and we have
introduced the differential operator
%%%
\begin{equation}
 \hat{\cal D}=\frac{d^2}{d r_*^2}+\omega^2-F\frac{\ell(\ell+1)}{r^2}\,.\label{D}
\end{equation}
%%%
where $r_*$ is the standard tortoise coordinate defined as
$dr/dr_*=(r-r_-)(r-r_+)/r^2$, and $r_\pm=M\pm\sqrt{M^2-Q^2}$ are the
outer ($r_+$) and Cauchy ($r_-$) horizons of a RN BH.  The functions
$Z_i^{-}$ and $Z_i^{+}$ are linear combinations of axial and polar
variables, respectively, and they are also combinations of
gravitational and electromagnetic perturbations. A step-by-step
derivation of these equations is presented in
Appendix~\ref{app:pertdeqs} and in a publicly available {\scshape
  Mathematica} notebook~\cite{webpage}. The axial potentials
$V^{(i,-)}$ and $W^{(i,-)}$ are listed in Appendix~\ref{app:axial},
while the polar potentials $V^{(i,+)}$ and $W^{(i,+)}$ are so lengthy
that we decided to make them available only through the {\scshape
  Mathematica} notebook \cite{webpage} in order to save space.

We have integrated the coupled system~\eqref{eF} and computed the
corresponding eigenfrequencies using two independent methods, which
are described in Section~\ref{sec:computingQNMs} (see also
Ref.~\cite{Pani:2013pma} for a review).
For any value of $Q$, our analysis allows us to extract the
first-order corrections to the complex QNM frequencies
$\omega=\omega_R+i\omega_I$:
\begin{eqnarray}
 \omega_R&=&\omega_R^{(0)}+\tilde{a}m\omega_R^{(1)}+{\cal O}(\tilde{a}^2),\label{wR}\\
 \omega_I&=&\omega_I^{(0)}+\tilde{a}m\omega_I^{(1)}+{\cal O}(\tilde{a}^2),\label{wI}
\end{eqnarray}
where $\omega_R^{(i)}$ and $\omega_I^{(i)}$ are functions of $Q$ and
of the multipolar index $\ell$, and the $m$-dependence has been
factored out. 

Section \ref{sec:results} presents our numerical results. We begin by
studying scalar perturbations of KN BHs, for which the perturbation
equations are separable and QNM frequencies can be computed
``exactly'' in the Teukolsky formalism \cite{Berti:2005eb}. By
comparing QNM frequencies in the slow-rotation approximation to the
Teukolsky-based results, we find that the relative error of the
slow-rotation approximation is less than 1\% as long as $J/J_{\rm
  max}\lesssim 0.3$, where $J_{\rm max}$ is the maximum allowed KN
spin for any fixed value of the electric charge.

Figure~\ref{fig:modes} shows our main numerical results for the
fundamental gravito-electromagnetic perturbations with $\ell=2,3$,
which are the most relevant for gravitational-wave emission (see
e.g.~\cite{Berti:2005ys}). In each panel we show four curves,
corresponding to the axial and polar ``gravitational'' and
``electromagnetic'' modes (as defined in the decoupled $Q=0$ limit:
see Section~\ref{sec:results} for details).  The zeroth-order terms
shown in the small left panels are simply RN QNMs; they agree with
continued-fraction solutions of the equations first derived by Zerilli
\cite{Zerilli:1974ai}, as computed by Leaver~\cite{Leaver:1990zz}.

Gravito-electromagnetic perturbations of nonspinning BHs in general
relativity have a noteworthy property that was proved by
Chandrasekhar~\cite{Chandra}: the polar and axial potentials can be
written in terms of a superpotential, which implies that the polar and
axial QNMs are \emph{isospectral}~\cite{Berti:2009kk}.
A priori, there is no reason why such a remarkable property should
hold true also for KN BHs. 

In addition to computing the QNMs of a KN BH in a full consistent
setting for the first time, perhaps the most important result of our
numerical study is strong evidence that the axial and polar sectors of
KN gravito-electromagnetic perturbations are indeed isospectral to
first order in the BH spin. In the inset of Fig.~\ref{fig:modes} we
show the relative difference between the coefficients of axial and
polar modes as functions of $Q$: our results are consistent with
isospectrality within the numerical errors implicit in our
method. This is further discussed in Section~\ref{sec:results}, where
we provide additional evidence that higher multipoles (with $\ell>2$)
and higher overtones (for a given $\ell$) are also isospectral to
first order in rotation.

These numerical results lead us to the tantalizing conjecture that the
modes of a KN BH may be isospectral to \emph{any order} in
$\tilde{a}$.  We conclude our paper in Section~\ref{sec:conclusions}
with a brief discussion of this conjecture and of possible ways to put
it to the test (see also~\cite{Pani:2013ija}).

%%%%%%
\section{Formalism}~\label{sec:formalism}
%%%%%%
Let us consider Einstein-Maxwell theory:
%%%
\begin{equation}
 S=\int d^4x\sqrt{-g}\left(R-F_{\mu\nu}F^{\mu\nu}\right)\,,
\end{equation}
where $R$ is the Ricci scalar and $F_{\mu\nu}=\partial_\mu
A_\nu-\partial_\nu A_\mu$ is the Maxwell field strength. The KN metric~\eqref{KNmetric}
is the most general stationary electrovacuum solution of this
theory. Here and in the following we linearize in the
spin parameter $\tilde a\equiv J/M^2$, neglecting terms of order
${\cal O}(\tilde a^2)$. To this order, the KN metric reads
%%%
\begin{align}
ds^2_0=g_{\mu\nu}^{(0)}dx^\mu dx^\nu=&-F(r)dt^2 +F(r)^{-1}dr^2+r^2d^2\Omega\nn\\
&-2\varpi(r)\sin^2\th d\varphi dt\,,
\label{metric0}
\end{align}
%%%%
where
%%%
\begin{eqnarray}
  F(r)&=&1-\frac{2M}{r}+\frac{Q^2}{r^2},\\
\varpi(r)&=&\frac{2\tilde{a} M^2}{r}-\frac{\tilde{a}Q^2M}{r^2},
\end{eqnarray}
%%%%
and the background electromagnetic potential is given by
%%%%%%
\begin{equation}
A_\mu=\left(\frac{Q}{r},0,0,-\frac{\tilde{a}Q M}{r}\sin^2\vartheta\right)\,.
\end{equation}
The magnetic field is given by the curl of the three-potential above. Thus, the
presence of \emph{both} rotation and charge ($\tilde a Q\neq0$) induces a
magnetic field in the $(\vartheta,\varphi)$ directions.
\subsection{Harmonic decomposition}
%%%%

We linearize the metric as
%%%
\begin{equation}
 g_{\mu\nu}=g_{\mu\nu}^{(0)}+h_{\mu\nu},
\end{equation}
and we decompose the metric perturbations $h_{\mu\nu}$ in the 
Regge--Wheeler gauge:
\begin{equation}
h_{\mu\nu}=\left(\begin{array}{cc|cc}
H_0^{{\ell}}Y^{{\ell}}&H_1^{{\ell}}Y^{{\ell}}&h_0^{{\ell}}S_\vartheta^{{\ell}}
&h_0^{{\ell}}S_\varphi^{{\ell}}\\
H_1^{{\ell}}Y^{{\ell}}&H_2^{{\ell}}Y^{{\ell}}&h_1^{{\ell}}S_\vartheta^{{\ell}}
&h_1^{{\ell}}S_\varphi^{{\ell}}\\\hline
h_0^{{\ell}}S_\vartheta^{{\ell}}&h_1^{{\ell}}S_\vartheta^{{\ell}}&r^2K^{{\ell}}Y^{{\ell}}&0\\
h_0^{{\ell}}S_\varphi^{{\ell}}&h_1^{{\ell}}S_\varphi^{{\ell}}&0&r^2K^{{\ell}}\sin^2\vartheta Y^{{\ell}}\\
\end{array}\right),\label{expmetric}
\end{equation}
where $Y^{{\ell}}=Y^{{\ell}}(\vartheta,\varphi)$ are the ordinary
scalar spherical harmonics,
$(S_\vartheta^{{\ell}},S_\varphi^{{\ell}})\equiv
\left(-{Y^{{\ell}}_{,\varphi}}/{\sin\vartheta},\sin\vartheta
Y^{{\ell}}_{,\vartheta}\right)$ are the axial vector harmonics, and
$H_{0,1,2}^{{\ell}},\,h_{0,1}^{{\ell}},\,K^{{\ell}}$ are functions of
$(t,\,r)$. Here and in the following, a sum over the harmonic indices
$\ell$ and $m$ (with $|m|<\ell$) is implicit. We will append the
relevant multipolar index $\ell$ to any perturbation variable but omit
the index $m$, because in an axisymmetric background it is possible to
decouple the perturbation equations so that all quantities have the
same value of $m$.

We expand the electromagnetic potential as follows \cite{Rosa:2011my}:
\begin{equation}\label{expansion_maxwell}
\delta A_{\mu}(t,r,\vartheta,\varphi)= \left[
 \begin{array}{c} 0 \\ 0\\
u_{(4)}^{{\ell}} S_b^{{\ell}}/\Lambda\\
 \end{array}\right]+\left[ \begin{array}{c}u_{(1)}^{{\ell}} Y^{{\ell}}/r\\u_{(2)}^{{\ell}} Y^{{\ell}}/(r F) \\
 u_{(3)}^{{\ell}} Y_b^{{\ell}}/\Lambda\\ \end{array}\right]\,,
\end{equation}
where $\Lambda=\ell(\ell+1)$, $b=(\vartheta,\varphi)$,
$Y_b^\ell=(Y_{,\vartheta}^\ell,Y_{,\varphi}^\ell)$ are the polar vector
harmonics, and $u_{(1,2,3,4)}^{{\ell}}$ are functions of $(t,\,r)$.
Inserting the harmonic expansion of the metric
perturbations~\eqref{expmetric} and of the Maxwell
field~\eqref{expansion_maxwell} into the linearized Einstein equations
we find the equations for the perturbation functions to linear order
in $\tilde a$. The latter naturally separate into three
groups~\cite{Kojima:1992ie,Pani:2012vp,Pani:2012bp,Pani:2013pma}.  By
denoting the linearized Einstein equations as $\delta {\cal
  E}_{\mu\nu}=0$, the first group is formed by the equations
$\delta{\cal E}_{(I)}=0$, where $I=0,1,2,3$ is shorthand notation for
$(tt),\,(tr),\,(rr)$ and $(+)$, respectively, and we have defined
$\delta{\cal E}_{(+)}\equiv\delta{\cal E}_{\vartheta\vartheta}
+{\delta {\cal E}_{\varphi\varphi}}/{\sin^2\vartheta}$. The equations
can be cast in the form
\begin{equation}
\delta{\cal E}_{(I)}\equiv (A^{(I)}_{{\ell}}+{\tilde A}^{(I)}_{{\ell}}\cos\th)Y^{{\ell}}
+B^{(I)}_{{\ell}}\sin\th Y^{{\ell}}_{,\vartheta}+C^{(I)}_{{\ell}} Y^{{\ell}}_{,\varphi}=0\label{eqG1}
\end{equation}
where $A^{(I)}_{{\ell}}\,,{\tilde
  A}^{(I)}_{{\ell}}\,,B^{(I)}_{{\ell}}\,,C^{(I)}_{{\ell}}$ are
combinations of the perturbation functions
$\{H_{0,1,2}^{{\ell}},\,h_{0,1}^{{\ell}},\,K^{{\ell}},\,u_{(1,2,3,4)}^{{\ell}}\}$.
%%%
The second group is formed by the equations $\delta{\cal E}_{L\vartheta}=0$,
with $L=0,1$ corresponding to the $t,r$ components, respectively. They can be
cast in the form
%%%
\begin{eqnarray}
&\delta{\cal E}_{(L\vartheta)}&\equiv(\alpha^{(L)}_{{\ell}}+{\tilde \alpha}^{(L)}_{{\ell}}\cos\th) Y^{{\ell}}_{,\vartheta}-
(\beta^{(L)}_{{\ell}}+{\tilde \beta}^{(L)}_{{\ell}}\cos\th)\frac{ Y^{{\ell}}_{,\varphi}}{\sin\th}\nn\\
&&+\eta^{(L)}_{{\ell}}\sin\th Y^{{\ell}}+\xi^{(L)}_{{\ell}}X^{{\ell}}+
\chi^{(L)}_{{\ell}}\sin\th W^{{\ell}}=0,\label{eqG2a}\\
%%%%
&\delta{\cal E}_{(L\varphi)}&\equiv(\beta^{(L)}_{{\ell}}+{\tilde \beta}^{(L)}_{{\ell}}\cos\th) Y^{{\ell}}_{,\vartheta}+
(\alpha^{(L)}_{{\ell}}+{\tilde \alpha}^{(L)}_{{\ell}}\cos\th)\frac{ Y^{{\ell}}_{,\varphi}}{\sin\th}\nn\\
&&+\zeta^{(L)}_{{\ell}}\sin\th Y^{{\ell}}+\chi^{(L)}_{{\ell}}X^{{\ell}}-
\xi^{(L)}_{{\ell}}\sin\th W^{{\ell}}=0,\label{eqG2b}
\end{eqnarray}
where $\alpha^{(L)}_{{\ell}}\,,{\tilde \alpha}^{(L)}_{{\ell}}\,,\beta^{(L)}_{{\ell}}\,,{\tilde \beta}^{(L)}_{{\ell}}\,,
\eta^{(L)}_{{\ell}}\,,\xi^{(L)}_{{\ell}}\,,\zeta^{(L)}_{{\ell}}\,,\chi^{(L)}_{{\ell}}$
are combinations of the perturbation functions.
The third group is (defining $\delta{\cal E}_{(-)}\equiv\delta{\cal
  E}_{\vartheta\vartheta} -{\delta {\cal
    E}_{\varphi\varphi}}/{\sin^2\vartheta}$)
\begin{eqnarray}
\delta {\cal E}_{(\vartheta\varphi)}&\equiv&	
f_{{\ell}}\sin\th\pa_{\th} Y^{{\ell}}+g_{{\ell}} Y^{{\ell}}_{,\varphi}+s_{{\ell}}
\frac{X^{{\ell}}}{\sin\th}+t_{{\ell}}W^{{\ell}}=0,\nn\\ \label{eqG3a}\\
\delta {\cal E}_{(-)}&\equiv&
g_{{\ell}}\sin\th\pa_{\th} Y^{{\ell}}-f_{{\ell}} Y^{{\ell}}_{,\varphi}-t_{{\ell}}
\frac{X^{{\ell}}}{\sin\th}+s_{{\ell}}W^{{\ell}}=0\,,\nn\\ \label{eqG3b}
\end{eqnarray}
%%%%%
where we have defined 
\beq
X^{{\ell}}&\equiv&2(Y^{{\ell}}_{,\th\ph}-\cot\th Y^{{\ell}}_{,\ph})\,,\\
W^{{\ell}}&\equiv&Y^{{\ell}}_{,\th\th}-\cot\th Y^{{\ell}}_{,\th}-\frac{Y^{{\ell}}_{,\ph\ph}}{\sin^2\th}\,.
\eeq
%
% \end{widetext}
%%%
The linearized Maxwell equations can be also recast in a similar
form. The $t-$ and $r-$components belong to the first group and can be
arranged in the form of Eq.~\eqref{eqG1} with $I=4,5$,
respectively. The $\vartheta-$ and $\varphi-$ components belong to the
second group, and can be written as in Eq.~\eqref{eqG2a}
and~\eqref{eqG2b} with $L=2$.
%%%
The coefficients can be split into two sets:
 \begin{eqnarray}
 \text{Polar:}\qquad && A^{(I)}_{{\ell}},\quad C^{(I)}_{{\ell}},\quad \alpha^{(L)}_{{\ell}},\quad \tilde\beta^{(L)}_{{\ell}},\nn\\
 &&\zeta^{(L)}_{{\ell}},\quad \xi^{(L)}_{{\ell}},\quad s_{{\ell}},\quad f_{{\ell}},\nn\\
%%%
 \text{Axial:}\qquad &&\tilde A^{(I)}_{{\ell}},\quad B^{(I)}_{{\ell}},\quad \beta^{(L)}_{{\ell}},\quad \tilde\alpha^{(L)}_{{\ell}},\nn\\
 &&\eta^{(L)}_{{\ell}},\quad \chi^{(L)}_{{\ell}},\quad t_{{\ell}},\quad g_{{\ell}},\nn
\end{eqnarray}
%%%
% \end{widetext}
%%
where henceforth the indices $I=0,...,5$ and $L=0,1,2$ account for both the
Einstein and Maxwell equations. The explicit form of the coefficients is not
particularly illuminating, and it is given in a publicly available {\scshape
  Mathematica} notebook~\cite{webpage}. The crucial point is to recognize that
the coefficients above are purely radial functions, i.e. the entire angular
dependence has been factored out from the field equations.
%%%%
\subsection{Separation of the angular dependence}
%%%

The separation of the angular dependence of Einstein's equations for a
slowly-rotating star was performed in Ref.~\cite{Kojima:1992ie}. The
procedure has been extended to general slowly rotating BH solutions in
Refs.~\cite{Pani:2012vp,Pani:2012bp} (see also~\cite{Pani:2013pma}); we
refer the reader to those papers for details. Using the orthogonality
properties of scalar spherical harmonics, from Eq.~\eqref{eqG1} we get
\begin{eqnarray}
&&A^{(I)}_{{\ell}}+{i} mC^{(I)}_{{\ell}}+\cQ_{{\ell}}\left[{\tilde A}^{(I)}_{{\ell-1}}
+(\ell-1){B}^{(I)}_{{\ell-1}}\right]+\nn\\
&&\cQ_{{\ell+1}}\left[{\tilde A}^{(I)}_{{\ell+1}}
-(\ell+2){ B}^{(I)}_{{\ell+1}}\right]=0,\label{decG1}
\end{eqnarray}
%%%
where we have defined ${\cal Q}_{{\ell}}^2=(\ell^2-m^2)/(4\ell^2-1)$.
From the orthogonality of vector spherical harmonics and using
Eqs.~\eqref{eqG2a}-\eqref{eqG2b} we get
\begin{eqnarray}
 &&\Lambda \alpha^{(L)}_{{\ell}}+{i} m\left[(\ell-1)(\ell+2)\xi^{(L)}_{{\ell}}
-{\tilde\beta}^{(L)}_{{\ell}}-\zeta^{(L)}_{{\ell}}\right]+\nn\\
&&\cQ_{{\ell}}(\ell+1)\left[(\ell-2)(\ell-1)\chi^{(L)}_{{\ell-1}}
+(\ell-1){\tilde\alpha}^{(L)}_{{\ell-1}}-\eta^{(L)}_{{\ell-1}}\right]-\nn\\
&&\cQ_{{\ell+1}}\ell\left[(\ell+2)(\ell+3)\chi^{(L)}_{{\ell+1}}
-(\ell+2){\tilde\alpha}^{(L)}_{{\ell+1}}-\eta^{(L)}_{{\ell+1}}\right]=0,\nn\\
\label{decG2a}\\
%&&\nn\\
&&\Lambda \beta^{(L)}_{{\ell}}+{i} m\left[(\ell-1)(\ell+2)\chi^{(L)}_{{\ell}}
+{\tilde\alpha}^{(L)}_{{\ell}}+\eta^{(L)}_{{\ell}}\right]-\nn\\
&&\cQ_{{\ell}}(\ell+1)\left[(\ell-2)(\ell-1)\xi^{(L)}_{{\ell-1}}-
(\ell-1){\tilde\beta}^{(L)}_{{\ell-1}}+\zeta^{(L)}_{{\ell-1}}\right]+\nn\\
&&\cQ_{{\ell+1}}\ell\left[(\ell+2)(\ell+3)\xi^{(L)}_{{\ell+1}}
+(\ell+2){\tilde\beta}^{(L)}_{{\ell+1}}+\zeta^{(L)}_{{\ell+1}}\right]=0\,.\nn\\
\label{decG2b}
\end{eqnarray}
%%%%
% where $L=0,1,2$.
%
Finally, the orthogonality of tensor spherical harmonics applied to
Eqs.~\eqref{eqG3a} and~\eqref{eqG3b} yields
%%%
\begin{eqnarray}
&&0=\Lambda s_{{\ell}}-{i} m f_{{\ell}}-\cQ_{{\ell}}(\ell+1)g_{{\ell-1}}+\cQ_{{\ell+1}}\ell g_{{\ell+1}} \label{decG3a}\,,\\
&&0=\Lambda t_{{\ell}}+{i} m g_{{\ell}}-\cQ_{{\ell}}(\ell+1)f_{{\ell-1}}+\cQ_{{\ell+1}}\ell f_{{\ell+1}}\,. \label{decG3b}
\end{eqnarray}
%%%
Summarizing, once we fix the value of $m$, truncating the expansion in
$\ell$ to a value $\ell_{\rm max}$, the index $\ell$ can have
$n_\ell=\ell_{\rm max}-|m|+1$ possible values; our separation
procedure in the slow-rotation limit yields a system of $14n_\ell$
coupled, ordinary differential equations (10 for the gravitational
sector and 4 for the Maxwell sector for each $\ell$), given in
Eqs.~\eqref{decG1}, \eqref{decG2a}-\eqref{decG2b} and
\eqref{decG3a}-\eqref{decG3b}. Their explicit form is available
online~\cite{webpage}.

%%%%%%%%%%%%%%%%%%%%%
\subsection{First-order corrections to the eigenvalue equations}
%%%%%%%%%%%%%%%%%%%%%
In the case of slow rotation ($\tilde a\ll1$), a Laporte-like
selection rule couples perturbations with harmonic index $\ell$ with
those having opposite parity and harmonic index
$\ell\pm1$~\cite{ChandraFerrari91,Kojima:1992ie,Pani:2012vp,Pani:2012bp}. This
is a consequence of using a basis of spherical harmonics in a
nonspherical background.
However, as discussed in detail
in~\cite{1993ApJ...414..247K,1993PThPh..90..977K,Pani:2012bp}, the
couplings to the $\ell\pm1$ terms do not contribute to the QNM
spectrum to first order in $\tilde a$. For this reason, we shall
neglect these terms in the following.
%%%%
In this way we get a system of $14$ coupled differential equations for
any chosen value of $\ell$.

Once the couplings among different $\ell$'s are neglected, we can
simply fix a value of $\ell$ (and in order to simplify the notation we
will drop the index $\ell$ from all perturbation functions). We
perform a Fourier decomposition by assuming that all perturbations
have a time dependence $\sim e^{-i\omega t}$. 

As we explicitly show in Appendix~\ref{app:pertdeqs}, axial and polar
perturbations decouple, and Eqs.~\eqref{decG1}-\eqref{decG3b} can be
reduced to
two coupled, second-order differential equations (one pair for each
parity), that we have already written down in the Introduction
[Eq.~\eqref{eF}].
These equations display the same symmetries as the master equations
for a RN BH~\cite{Chandra}, and indeed they exactly reduce to those
equations in the nonrotating case. We remark in passing that the DF
equations previously used to investigate gravito-electromagnetic
perturbations of the KN metric do not satisfy this 
requirement. Eqs.~\eqref{eF} are the main result of this paper. They
contain two first-order corrections in the rotational parameter
$\tilde{a}$. The first term couples the functions $Z_i^\pm$ with the
same $Z_i^\pm$. It is responsible for a Zeeman-like splitting of the
eigenfrequencies, which breaks the degeneracy in the azimuthal index
$m$. The second line in Eqs.~\eqref{eF} is more interesting. First, it
is proportional to the combination $\tilde{a}Q^2$, so it is vanishing
when the BH is nonspinning or uncharged. Furthermore, this term
couples the function $Z_1^+$ with the function $Z_2^+$, and the
function $Z_1^-$ with the function $Z_2^-$.

We remark that this coupling disappears when $\tilde a=0$, but this does not
mean that gravitational and electromagnetic perturbations decouple in that
limit: whenever $Q\neq0$, the functions $Z_j^\pm$ are combinations of
electromagnetic and gravitational perturbations, which are then coupled.
Electromagnetic and gravitational perturbations only
decouple when $Q=0$ (e.g., for Schwarzschild or Kerr BHs) because the functions
$Z_1^\pm$ describe pure gravitational perturbations, the functions $Z_2^\pm$
describe pure electromagnetic perturbations, and the coupling term in~\eqref{eF}
vanishes (see e.g. Ref.~\cite{Chandra}).

Despite the complicated form of Eqs.~\eqref{eF}, their asymptotic
behavior is simple. The asymptotic solutions at the horizon and at
infinity are plane waves, namely:
\begin{eqnarray}
 Z_j^\pm(r)\sim \left\{\begin{array}{l}
                        e^{{i}\omega r_*},\qquad \hspace{1.35cm} r\to\infty\\
                        e^{-{i} (\omega-m\Omega_H) r_*},\qquad r\to r_+
                       \end{array}\right.\,.
\label{asymp_hor}
\end{eqnarray}
As shown by the behavior above, near the BH horizon we obtain the typical frame-dragging effect: a static observer at infinity would see a
wave whose wave number is
\begin{equation}
k_H=\omega-m\Omega_H\sim\omega-\frac{\tilde{a}m}{M (1+{\tilde a}_{\rm max})^2}+{\cal O}(\tilde{a}^3),\label{kH}
\end{equation}
where we have assumed $\omega M\ll \tilde{a}$, and 
\begin{equation}
\tilde a_{\rm max}\equiv J_{\rm max}/M^2=\sqrt{1-(Q/M)^2} \label{amax}
\end{equation}
is the maximum spin parameter of a KN BH and $\Omega_H=-\lim_{r\to
  r_+}g_{t\varphi}/g_{\varphi\varphi}$ is the angular velocity at the
horizon of locally nonrotating observers to first order in
$\tilde{a}$. In the small-charge limit,
$k_H\sim\omega-\tilde{a}m/(4M)-\tilde{a}mQ^2/(8M^3)+{\cal
  O}[(Q/M)^4]$.

According to Eq.~\eqref{asymp_hor}, if $\omega<m\Omega_H$ an observer
at infinity would see waves coming out of the horizon. This
corresponds to extraction of energy from a spinning BH, resulting in
superradiant amplification of the
wave~\cite{Teukolsky:1974yv}. Therefore it is not necessary to solve
the linearized field equations in order to show that superradiance in
a KN spacetime occurs also for gravito-electromagnetic perturbations,
similarly to the scalar case. A note of caution is necessary in this
regard: as discussed in Ref.~\cite{Pani:2012bp}, we must include
\emph{second-order} terms in the expansion for a consistent treatment
of superradiance in the slow-rotation approximation. The reason is
that the superradiance condition implies $\omega<\Omega_H\sim{\cal
  O}(\tilde a)$, 
therefore the $\omega^2$ term in the wave equation becomes of the 
same order as ${\cal O}({\tilde a}^2)$ terms.
%and the energy of the wave scales like $\omega^2\sim{\cal O}(\tilde a^2)$. 
However the results of
Ref.~\cite{Pani:2012bp} show that (at least in some specific cases)
first- and second-order results are in qualitative (and sometimes in
remarkably good quantitative) agreement even in the superradiant
regime.

%%%%%%%%%%%%%%%%%%
\section{Quasinormal mode calculation for coupled systems}\label{sec:computingQNMs}
%%%%%%%%%%%%%%%%%%
After imposing physically-motivated boundary conditions at the horizon
and at infinity, Eqs.~\eqref{eF} form an eigenvalue problem for the
frequency $\omega$. Robust numerical methods to solve this class of
coupled eigenvalue problems have recently been extended to BH
spacetimes (cf.~\cite{Pani:2013pma} for a review).

We have solved the field equations~\eqref{eF} by two independent
techniques: a matrix-valued continued-fraction method and direct
integration~\cite{Rosa:2011my,Pani:2012bp}.

\subsection{Matrix-valued continued fractions}

In order to reduce Eqs.~\eqref{eF} to a matrix-valued recurrence relation, we
use the ansatz:
%%%
\begin{eqnarray}
 Z_j^\pm(r)&=&\frac{r_+(r_+-r_-)^{-4{i} M\omega-1}}{r}e^{-2{i}\omega r_+}(r-r_-)^{1+2{i} M\omega}\nn\\
&&e^{{i}\omega r} z^{\frac{-{i} k_H r_+^2}{r_+-r_-}}\sum_n a_n^{\pm,j}z^n, \label{ansatzCF}
\end{eqnarray}
%%%
where $z=(r-r_+)/(r-r_-)$ and the $\pm$ superscripts (not to be
confused with the subscripts of the inner and outer horizon, $r_\pm$)
refer to the  polar case [``plus'' superscript in Eqs.~\eqref{eF}] and
to the axial case [``minus'' superscript in Eqs.~\eqref{eF}],
respectively. In the axial case we obtain a ten-term matrix-valued
recurrence relation, whereas in the polar case we obtain a twelve-term
matrix-valued recurrence relation. They can be reduced to 
three-term recurrence relations using a matrix analog of Gaussian
elimination~\cite{Pani:2013pma}. Finally, the problem of finding the
QNMs of the system is reduced to finding the complex roots of a
determinant obtained from a nested series of inverted
matrices~\cite{Pani:2013pma}.

\subsection{Direct integration}

In alternative, a direct integration technique can be shown to work
very well, as long as the modes' real part is sufficiently larger than
the imaginary part. This condition is typically satisfied by the
fundamental mode and (to a lesser extent) by the first overtone. In
this direct integration approach, the system~\eqref{eF} is integrated
numerically from the horizon out to infinity. At the horizon we set
the solution equal to some high-order series expansion of the form
\begin{equation}
  Z_j^\pm\sim e^{-i k_H r_*}\sum_n b_n^{\pm,j}(r-r_+)^n \qquad r\to r_+\,,\label{series_hor}
\end{equation}
where the coefficients $b_n^{\pm,j}$ $(n>0$, $j=1,2$) can be computed
in terms of $b_0^{\pm,i}$ by solving the near-horizon equations
order-by-order. Two independent solutions are obtained by choosing the
orthonormal bases $(b_0^{\pm,1},b_0^{\pm,2})=(1,0)$ and
$(b_0^{\pm,1},b_0^{\pm,2})=(0,1)$. At infinity the generic behavior of
the solutions reads
\begin{equation}
  Z_j^\pm\sim B_j^\pm e^{i \omega r_*}\sum_n \frac{c_n^{\pm,j}}{r^n}+
C_j^\pm e^{-i \omega r_*}\sum_n \frac{d_n^{\pm,j}}{r^n} \qquad r\to \infty\,.
\label{series_inf}
\end{equation}
The QNM boundary conditions correspond to imposing $C_j^\pm=0$, i.e.,
purely outgoing waves at infinity. The eigenfrequencies are computed
as the complex roots of (see e.g. \cite{Ferrari:2007rc})
%%%
\begin{equation}
 {\rm det}(\mathbf{S}^\pm)\equiv {\rm det}\begin{pmatrix}
C_1^{\pm,1} & C_1^{\pm,2}   \\ 
C_2^{\pm,1} & C_2^{\pm,2}   \\
\end{pmatrix}\,,
\end{equation}
%%%%
where the superscripts denote a particular vector of the basis,
i.e. $C_j^{\pm,1}$ are related to $(b_0^{\pm,1},b_0^{\pm,2})=(1,0)$
and $C_j^{\pm,2}$ are related to
$(a_0^{\pm,1},a_0^{\pm,2})=(0,1)$. For any given frequency we
integrate the system~\eqref{eF} twice and construct the matrix
$\mathbf{S}^\pm$. The QNMs are then obtained by imposing ${\rm
  det}(\mathbf{S}^\pm)=0$, which can be achieved via a shooting
method~\cite{Pani:2013pma}.

%%%%%%%%%%%%%%%%%%%%%%%%%%%
\section{Numerical results}\label{sec:results}
%%%%%%%%%%%%%%%%%%%%%%%%%%%

%%%%%%%%%%%%%%%%%
\subsection{Scalar Quasinormal Modes in the Slow-Rotation Limit}\label{scalartest}
%%%%%%%%%%%%%%%%%

As a test of the slow-rotation approximation, we have computed the
scalar QNMs of a KN BH to first order in $\tilde{a}$. Since these
modes can be computed exactly in the Teukolsky formalism (e.g. via
continued fractions~\cite{Berti:2005eb}), we can use these ``exact''
results to estimate the errors introduced by the slow-rotation
approximation. For any stationary and axisymmetric spacetime, the
scalar modes at first order in the angular momentum are governed by a
master equation~\cite{Pani:2012bp}. In the special case of a
background given by Eq.~\eqref{metric0}, the master equation reduces
to
\begin{equation}
 \frac{d^2\psi}{dr_*^2}+\left[\omega^2-\frac{2m\omega\varpi(r)}{r^2}
-F\left(\frac{\ell(\ell+1)}{r^2}+\frac{F'}{r}\right)\right]\psi=0\,.
\label{scalar}
\end{equation}
The corresponding eigenvalue problem can be solved with standard
continued-fraction
techniques~\cite{Leaver:1990zz,Berti:2004md,Berti:2005eb}. Using the
ansatz~\eqref{ansatzCF} for $\psi$ yields a five-term recurrence
relation, whose coefficients are listed in Appendix~\ref{app:CFs}. The
five-term recurrence relation can be reduced to a three-term relation
via standard Gaussian elimination.

In Ref.~\cite{Pani:2013ija} we computed the relative error of the
slow-rotation approximation with respect to the exact result. We found
a near-universal behavior of the percentage errors as functions of $J/J_{\rm max}=\tilde a/{\tilde a}_{\rm max}$, where ${\tilde a}_{\rm max}$ is the maximum allowed value of the spin
parameter [corresponding to an extremal KN BH with the given charge
  $Q$, cf. Eq.~\eqref{amax}] for all values
of $Q$. 
%%%%%%
This near-universality suggests that the parameter ${\tilde
  a}_{\rm max}$ [which appears explicitly in the QNM boundary
  conditions~\eqref{kH}] plays a fundamental role in our perturbative
scheme (see also the discussion in Section \ref{GEmKN}). In other
words, the slow-rotation
approximation is accurate {\em only far from extremality}, i.e. when
\begin{equation}
\tilde a\ll{\tilde a}_{\rm max}=\sqrt{1-\left(\frac{Q}{M}\right)^2}<1\,.
\end{equation}
This condition is obviously stronger than the condition one may expect
to apply a priori, i.e. $\tilde a\ll1$, and it implies that the
slow-rotation approximation will break down even for small values of
the spin as $Q\to M$. A posteriori, this result is not too surprising,
and it is consistent with the requirement that the perturbation
parameter $\tilde{a}$ must be smaller than {\em any dimensionless
  quantity} characterizing the background spacetime.

According to Ref.~\cite{Pani:2013ija}, the slow-rotation
approximation is accurate within one percent as long as $J/J_{\rm
  max}\lesssim 0.3$, and it is still accurate within $3\%$ for
$J/J_{\rm max}\lesssim 0.5$.
%

%%%%%%%%%%%%%%%%%
\subsection{Gravito-Electromagnetic Quasinormal Modes at First Order in Rotation
}\label{GEmKN}
%%%%%%%%%%%%%%%%%

We have computed the fundamental mode and the first overtone of the
gravito-electromagnetic QNMs of a KN BH in the axial and polar sectors
for $\ell=2$ and $\ell=3$. The numerical solution of both the axial
and polar perturbation equations~\eqref{eF} is extremely challenging,
because their explicit form is lengthy and complicated. 

We have computed the axial modes using both the matrix
continued-fraction method and direct integration. The results agree
extremely well, so that the two methods validate each other. In the
polar case the equations are too complicated to be cast in a tractable
continued-fraction form, even using algebraic manipulation software
like {\scshape Mathematica}. In this case the only viable technique
turned out to be direct integration. Even the task of numerical
integration is challenging due to the complexity of the field
equations, but reasonably accurate results can be obtained by using
high-order series expansions at the horizon and at infinity and by
requiring better numerical precision in the integration. The
difficulties we met in integrating the first-order equations suggest
that a second-order analysis, while possible in principle, may be very
challenging from the algebraic and computational standpoints.

For any value of $Q$, our analysis allows us to extract the
first-order corrections to the QNM spectrum, as defined in
Eqs.~\eqref{wR} and~\eqref{wI}.  At first order the $m$-dependence can
be factored out, so the calculation is complete once we know the
functions $\omega_R^{(1)}$ and $\omega_I^{(1)}$.  Furthermore, we remark 
that first-order corrections vanish identically when $m=0$.

We find four families of modes, two of which are associated to axial
perturbations, while the other two are associated to polar perturbations. 
In the $Q=0$ limit the two families with a given parity reduce to the
gravitational and electromagnetic modes of a Kerr BH. Therefore, with
a slight abuse of notation, we will dub the two families
``gravitational'' and ``electromagnetic'' also in the general case. It
should be stressed that, when $Q\neq0$, oscillations involving any of
these modes excite both electromagnetic and gravitational
perturbations.

We also mention that we have carried out a further check of our results. 
We have extracted ${\cal O}(\tilde a)$ corrections of the QNMs from a fit 
of the ``full'' numerical solution of the Teukolsky equations in the Kerr 
background, and we have verified that this procedure matches our results 
for $Q=0$ within a relative error $\lesssim 0.1\%$ for
$\omega_R^{(1)}(Q=0)$. The error is only a few times larger for
$\omega_I^{(1)}(Q=0)$.

%%%%%%%%%%%%%%%%%
\subsubsection{Isospectrality}
%%%%%%%%%%%%%%%%%
For generic values of $\ell$, gravito-electromagnetic perturbations of
nonspinning BHs in general relativity are
isospectral~\cite{Chandra,Berti:2009kk}. The left panels of
Fig.~\ref{fig:modes} are fully consistent with isospectrality within
our numerical accuracy, and this is a nontrivial consistency check of
our method\footnote{To the best of our knowledge, in the published
  literature there are no studies checking isospectrality for RN BHs
  by an explicit calculation of polar QNMs. This is an interesting
  by-product of our analysis.}.

A priori, there is no reason to expect that such a remarkable property
should hold true also for KN BHs. Isospectrality is easily broken: for
example it is well known that polar and axial modes are {\em not}
isospectral (even for nonrotating BHs) if the cosmological constant is
nonzero \cite{Mellor:1989ac,Cardoso:2001bb,Berti:2003ud}, if the
underlying theory is not general
relativity~\cite{Cardoso:2009pk,Molina:2010fb}, or in higher
dimensions~\cite{Berti:2009kk}.

Our numerics provide strong evidence that the gravito-electromagnetic
modes of KN BHs are isospectral to first order in the angular
momentum. This was shown in Fig.~\ref{fig:modes} for the fundamental
mode with $\ell=2$ and $\ell=3$.  In the insets of
Fig.~\ref{fig:modes} we show the relative difference between the
coefficients of axial and polar modes as functions of $Q$: our results
are consistent with isospectrality within the numerical accuracy of
the direct integration method.

We believe that deviations from isospectrality are of a purely
numerical nature, being almost entirely due to the intrinsic errors of
the direct integration to compute QNMs. This issue was discussed in
detail in Ref.~\cite{Pani:2013ija}.  Here we simply add the
observation that, as shown in Fig.~\ref{fig:modes}, the fundamental
mode with $\ell=3$ (for which the direct integration is more accurate)
is numerically closer to isospectrality than the fundamental mode with
$\ell=2$.  We have checked that deviations from isospectrality
decrease as $\ell$ grows.

%%%%%%%%%%%%%%%%%
\subsubsection{Fitting formulae at first order in rotation}
%%%%%%%%%%%%%%%%%

Polar and axial modes are the same to first order within our numerical
accuracy, but the equations for axial modes, which are listed
explicitly in Eq.~(\ref{eF}) and Appendix~\ref{app:axial} of this
paper, are much simpler. We carried out a more extensive QNM
calculation working in the axial case, where our results can be
verified using two independent methods. Due to isospectrality, these
results cover the whole QNM spectrum of slowly rotating KN BHs.

We found that the zeroth- and first-order terms in Eqs.~\eqref{wR} and \eqref{wI} shown in
Fig.~\ref{fig:modes} are well fitted by functions of
the form
\begin{equation}
 M\omega^{(0,1)}_{R,I}= f_0 + f_1 y + f_2 y^2 +f_3 y^3+f_4 y^4\,,
\label{fit}
 \end{equation}
where we have defined a parameter $y=1-{\tilde a}_{\rm
  max}=1-\sqrt{1-Q^2/M^2}$ which is in one-to-one correspondence with
$Q$ (such that $y\in [0,\,1]$ as $Q\in [0,\,M]$), but is better suited
for fitting. As discussed in Section \ref{scalartest}, ${\tilde
  a}_{\rm max}$ seems to be the most appropriate dimensionless
quantity to normalize our perturbative parameter; in some sense, the
parameter $y$ measures the ``distance from extremality'' of the KN
metric. The $\ell$- and $n$-dependent fitting coefficients $f_i$ of
the functions $\omega_R^{(0,1)}$ and $\omega_I^{(0,1)}$ for a selected
subset of gravitational and electromagnetic modes are listed in
Table~\ref{tab:fit}, which extends a similar Table in
\cite{Pani:2013ija}.
%
%%%%%%%%%
%
\begin{table}[htb]
 \begin{tabular}{cc|ccccc}
  &	($\ell$,$n$,$s$)	& $f_0$& $f_1$& $f_2$& $f_3$& $f_4$ \\
\hline
$\omega_R^{(0)}$ & (2,0,1) &  0.4576 &  0.2659 &  0.0118 &  0.1228 & -0.1382 \\
$\omega_R^{(1)}$ & (2,0,1) &  0.0712 &  0.0769 &  0.0596 &  0.0727 & -0.0216 \\
$\omega_I^{(0)}$ & (2,0,1) & -0.0950 & -0.0184 &  0.0137 &  0.0132 &  0.0107 \\
$\omega_I^{(1)}$ & (2,0,1) &  0.0007 &  0.0043 &  0.0060 & -0.0089 &  0.0366 \\
\hline
$\omega_R^{(0)}$ & (2,0,2) &  0.3737 &  0.0525 &  0.0607 & -0.0463 & -0.0070 \\
$\omega_R^{(1)}$ & (2,0,2) &  0.0628 &  0.0676 &  0.0209 &  0.0823 & -0.0810 \\
$\omega_I^{(0)}$ & (2,0,2) & -0.0890 & -0.0055 &  0.0024 &  0.0214 & -0.0084 \\
$\omega_I^{(1)}$ & (2,0,2) &  0.0010 &  0.0014 &  0.0091 &  0.0174 &  0.0145 \\
\hline
\hline
$\omega_R^{(0)}$ & (2,1,1) &  0.4365 &  0.2793 &  0.0125 &  0.1399 & -0.1637 \\
$\omega_R^{(1)}$ & (2,1,1) &  0.0780 &  0.0785 &  0.0588 &  0.0776 & -0.0277 \\
$\omega_I^{(0)}$ & (2,1,1) & -0.2907 & -0.0515 &  0.0438 &  0.0364 &  0.0363 \\
$\omega_I^{(1)}$ & (2,1,1) &  0.0043 &  0.0138 &  0.0164 & -0.0230 &  0.1062 \\
\hline
$\omega_R^{(0)}$ & (2,1,2) &  0.3467 &  0.0546 &  0.0709 & -0.0292 & -0.0433 \\
$\omega_R^{(1)}$ & (2,1,2) &  0.0717 &  0.0764 &  0.0020 &  0.1959 & -0.2213 \\
$\omega_I^{(0)}$ & (2,1,2) & -0.2739 & -0.0157 &  0.0099 &  0.0668 & -0.0239 \\
$\omega_I^{(1)}$ & (2,1,2) &  0.0065 &  0.0070 &  0.0360 &  0.0254 &  0.0905 \\
\hline
\hline
$\omega_R^{(0)}$ & (3,0,1) &  0.6569 &  0.3684 & -0.0820 &  0.2851 & -0.2574 \\
$\omega_R^{(1)}$ & (3,0,1) &  0.0726 &  0.0768 &  0.0595 &  0.0617 & -0.0259 \\
$\omega_I^{(0)}$ & (3,0,1) & -0.0956 & -0.0177 &  0.0178 &  0.0074 &  0.0106 \\
$\omega_I^{(1)}$ & (3,0,1) &  0.0002 &  0.0032 &  0.0027 & -0.0002 &  0.0216 \\
\hline
$\omega_R^{(0)}$ & (3,0,2) &  0.5994 &  0.0790 &  0.1734 & -0.2019 &  0.0700 \\
$\omega_R^{(1)}$ & (3,0,2) &  0.0673 &  0.0693 &  0.0211 &  0.0791 & -0.0677 \\
$\omega_I^{(0)}$ & (3,0,2) & -0.0927 & -0.0043 & -0.0013 &  0.0292 & -0.0130 \\
$\omega_I^{(1)}$ & (3,0,2) &  0.0006 &  0.0014 &  0.0084 &  0.0058 &  0.0122 \\
\hline
\hline
$\omega_R^{(0)}$ & (3,1,1) &  0.6418 &  0.3782 & -0.0863 &  0.3104 & -0.2875 \\
$\omega_R^{(1)}$ & (3,1,1) &  0.0760 &  0.0786 &  0.0511 &  0.0924 & -0.0576 \\
$\omega_I^{(0)}$ & (3,1,1) & -0.2897 & -0.0509 &  0.0508 &  0.0333 &  0.0222 \\
$\omega_I^{(1)}$ & (3,1,1) &  0.0015 &  0.0098 &  0.0096 & -0.0050 &  0.0693 \\
\hline
$\omega_R^{(0)}$ & (3,1,2) &  0.5826 &  0.0819 &  0.1752 & -0.1753 &  0.0304 \\
$\omega_R^{(1)}$ & (3,1,2) &  0.0713 &  0.0736 &  0.0108 &  0.1287 & -0.1282 \\
$\omega_I^{(0)}$ & (3,1,2) & -0.2813 & -0.0123 & -0.0050 &  0.0972 & -0.0472 \\
$\omega_I^{(1)}$ & (3,1,2) &  0.0029 &  0.0059 &  0.0194 &  0.0374 &  0.0231 \\
\hline
\hline

\end{tabular}
\caption{Coefficients of the fit~\eqref{fit} for the real and
  imaginary part of a selected subset of gravito-electromagnetic
  modes. The values $(\ell,n)$ correspond to the multipolar index and
  the overtone number, respectively. Fundamental modes correspond to
  $n=0$. We denote by $s=1$ and $s=2$ the modes that in the decoupled
  $Q\to0$ limit are electromagnetic and gravitational in the
  Kerr background, respectively.  The fits~\eqref{fit}
  reproduce the data to within $1\%$ for $\omega_I^{(1)}$ and to
  within $0.1\%$ for the other quantities for any $Q\lesssim0.95 M$.
\label{tab:fit}}
\end{table} 

Our QNM calculations in the slow-rotation approximation can be seen as
an empirical confirmation of the stability of the KN metric. We have
looked for unstable modes in the region $0<Q<M$, $J\ll J_{\rm max}$
and for $\ell=2,3,4$ and we found none. This confirms early arguments
by Mashhoon, who used calculations in the eikonal limit to make a case
for the stability of the KN metric~\cite{Mashhoon:1985}. Notice
however that Mashhoon's results apply only to perturbation modes with
$\ell\gg 1$ and they rely on a geodesic analogy, rather than on a
self-consistent treatment of the perturbation equations. In this
sense, our findings provide the first self-consistent stability
analysis of the KN metric.
%%%%

%%%%%%%%%%%%%%%
\subsection{Comparison with the quasinormal modes of the Dudley-Finley equation}
\begin{figure}[thb]
\begin{center}
\begin{tabular}{c}
\epsfig{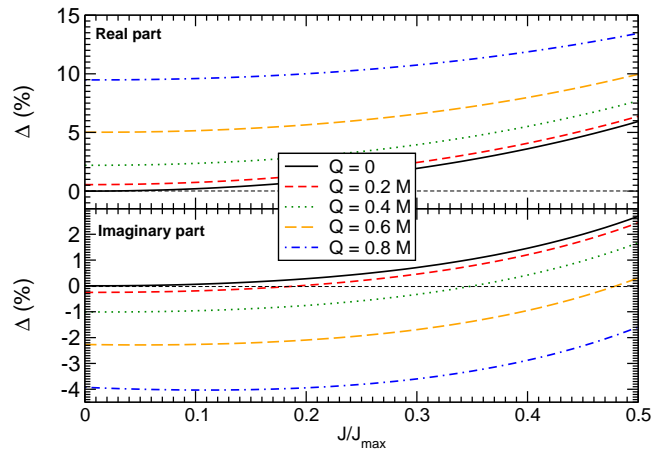}
\end{tabular}
\caption{Percentage deviation of QNM frequencies in the slow-rotation
  approximation with respect to the DF equation, for the $\ell=m=2$
  fundamental gravitational mode. As expected, the two approximations
  agree with each other when $Q\to0$, but they deviate from each other
  when $\tilde{a}\to0$. The discrepancy between the two approximations
  is nearly constant as long as $J\ll J_{\rm max}$, confirming
  that the DF equation is not very accurate in that regime when
  $Q\neq0$ (cf.~\cite{Berti:2005eb}). \label{fig:DF}}
\end{center}
\end{figure}

%%%%%%%%%%%%%%%
Finally, we can compare our results against the DF equation
\cite{Dudley:1977zz,Dudley:1978vd} to quantify the regime of validity
of both approximation schemes.  The results of this comparison for the
fundamental gravitational mode with $\ell=m=2$ are shown in
Fig.~\ref{fig:DF}.
%%%%

The DF equation reduces to the Teukolsky equation for Kerr BHs in the
limit $Q\to0$. However, it does \emph{not} reduce to the RN
Regge-Wheeler/Zerilli equations when $\tilde{a}\to0$
\cite{Berti:2005eb}. On the other hand, our slow-rotation
approximation is valid for any $Q$, but we must impose the condition
$J\ll J_{\rm max}$. In the region $J\ll J_{\rm max}$, the
slow-rotation approximation can be used to quantify the errors
introduced by the DF equation for any value of $Q$.

Figure~\ref{fig:DF} shows that the deviations between the DF and
slow-rotation calculations vanish when $Q\ll M$ and $J\ll J_{\rm
  max}$, i.e. in the region where the assumptions underlying both
approximations are consistent. As we increase $Q$ we observe an
increasing deviation of the DF modes with respect to the slow-rotation
calculation. The offset increases with $Q$, and it is nearly constant
for any $Q$ in the region $J\ll J_{\rm max}$: the errors introduced by
the DF approximation in the nonrotating case do not increase much when
(a small amount of) rotation is included. The deviations do increase
for larger values of $J$, but in that regime the slow-rotation
approximation is not reliable anymore.

%%%%%%%%%
\section{Conclusions}\label{sec:conclusions}
%%%%%%%%%
In this paper and the accompanying {\em Letter} \cite{Pani:2013ija} we
have presented the first fully-consistent analysis of the
gravito-electromagnetic QNMs of the KN metric. Working in a
slow-rotation approximation, the long-standing problem of
nonseparability of the perturbation equations can be evaded. We have
computed the gravito-electromagnetic QNMs to first order in the BH
spin and provided fitting formulae for the fundamental mode and first
overtone with $\ell=2,3$.

Furthermore, our numerical study of gravito-electromagnetic
perturbations shows strong numerical evidence for the isospectrality
of polar and axial gravito-electromagnetic perturbations of KN black
holes at linear order in rotation. It would be interesting to
understand whether isospectrality holds exactly, at all orders in
rotation.  An important extension of our work in this direction is to
include second-order effects. The causal structure of a spinning
metric starts differing from the nonspinning case at second order in
the angular momentum (e.g. changes in the horizon location and in the
ergoregion are of second order), thus if isospectrality holds true
also at second order, there is no fundamental reason to believe that
it is broken at higher order.

If isospectrality is an exact property of KN BHs, extensions of the
polar and axial field equations~\eqref{eF} to {\em any} order in the
spin should be related to each other by some transformation that
leaves the QNM spectrum invariant.  Even at linear order in rotation
isospectrality is a highly nontrivial property, in view of the mixing
of gravitational and electromagnetic perturbations. Hopefully our work
will stimulate further study to prove (or disprove) the conjecture
that isospectrality is an exact property of the KN spacetime.  This
conjecture may be verified using a brute-force extension of our work
to higher orders in rotation, numerical time evolutions along the
lines of \cite{Dorband:2006gg,Witek:2012tr,Dolan:2012yt}, or (ideally)
an analytical proof, perhaps similar to Chandrasekhar's proof in the
nonrotating case (see also Ref.~\cite{1999math.ph...9030L}).

In Ref.~\cite{Pani:2013ija} we have presented some possible extensions of 
our work.  If isospectrality turns out to be valid for any value of the angular 
momentum, an interesting avenue of research is to understand whether 
such property has implications in the context of the KN/CFT
conjecture~\cite{Hartman:2008pb,Hartman:2009nz}, which predicts that
the QNMs of the near-horizon KN geometry correspond to the poles of
the retarded Green's function of the dual chiral
CFT~\cite{Chen:2010i}.

Some further interesting applications concern nonasymptotically flat
spacetimes. Even for nonrotating RN~(anti-)de Sitter BHs
isospectrality is known to be partially broken, depending on the
relative size of the BH and (anti-)de Sitter horizon radii
\cite{Mellor:1989ac,Wang:2000gsa,Cardoso:2001bb,Berti:2003ud}. The
slow-rotation approximation may be used to understand whether similar
considerations also apply to Kerr-Newman (anti-)de Sitter. Indeed, our
approach can be easily extended to include a nonvanishing cosmological
constant. In the context of the AdS/CFT
correspondence~\cite{Maldacena:1997re}, the QNMs of a KN BH are dual
to thermal states of a CFT living in a rotating Einstein
universe~\cite{Caldarelli:1999xj,Hawking:1999dp}.

Finally, certain KN-AdS BHs embedded in $N=2$ four-dimensional
supergravity preserve half of the
supersymmetry~\cite{Kostelecky:1995ei,Caldarelli:1998hg}. This is
analogous to the case of asymptotically flat, extremal RN BHs, for which 
supersymmetry  implies a remarkable property of the QNMs: electromagnetic
perturbations with multipolar index $\ell$ are isospectral with
gravitational perturbations with index
$\ell+1$~\cite{Onozawa:1995vu}. Using this property, it is possible to
prove that the one-loop corrections to the BH entropy
cancel~\cite{Kallosh:1997ug}.  It would be interesting to understand
whether such supersymmetry implies a similar property for the
supersymmetric KN-AdS BH solutions found in
Refs.~\cite{Kostelecky:1995ei,Caldarelli:1998hg}. We note that such
solutions can be slowly rotating, so our framework can be directly
applied to this interesting problem.

We hope that our paper will stimulate further work in these, and
possibly other, directions.

%%%%%%%%%%%%%%%%%%%%%%%%%%%%%%%%%%%%%%%%%%%%%%%%%%%%%%%%%%%%%%%%%%%%%%%%%%%%%%
\begin{acknowledgments}
  We wish to thank Vitor Cardoso for useful discussions. This work was
  supported by the NRHEP--295189 FP7--PEOPLE--2011--IRSES Grant, and
  by FCT - Portugal through PTDC projects FIS/098025/2008,
  FIS/098032/2008, CERN/FP/123593/2011. E.B. was supported by NSF
  CAREER Grant No.~PHY-1055103. P.P. acknowledges financial support
  provided by the European Community through the Intra-European Marie
  Curie contract aStronGR-2011-298297.  Computations were performed on
  the ``Baltasar Sete-Sois'' cluster at IST, the cane cluster in
  Poland through PRACE DECI-7 ``Black hole dynamics in metric theories
  of gravity'', on Altamira in Cantabria through BSC grant
  AECT-2012-3-0012, on Caesaraugusta in Zaragoza through BSC grants
  AECT-2012-2-0014 and AECT-2012-3-0011, XSEDE clusters SDSC Trestles
  and NICS Kraken through NSF Grant~No.~PHY-090003, Finis Terrae
  through Grant CESGA-ICTS-234.
\end{acknowledgments}

%%%%%%%%%%%%%%%%%%%%%%%%%%%%%%%%%%%%%%%%%%%%%%%%%%%%%%%%%%%%%%%%%%%%%%%%%%%%%%
\appendix
%%%%%%%%%%%%%%%%%%%%%%%%%%%%%%%%%%%%%%%%%%%%%%%%%%%%%%%%%%%%%%%%%%%%%%%%%%%%%%

%%%%%%%%%%%%%%
\section{Derivation of the perturbation equations}\label{app:pertdeqs}
%%%%%%%%%%%%%
In this Appendix we derive the first-order equations~\eqref{eF} for
the axial and polar gravito-electromagnetic perturbations of a KN
BH. Various intermediate steps are presented in a supplementary
{\scshape Mathematica} notebook~\cite{webpage}.

%%%%%%%%%%%
\subsection{Axial sector}\label{app:axial}
%%%%%%%%%%%

As discussed in the main text and in Ref.~\cite{Pani:2012bp}, the
coupling between perturbations with different parity and different
harmonic index $\ell$ does not contribute to the QNM to first order in
$\tilde{a}$.  Neglecting the couplings to $\ell\pm1$ terms, the axial
sector is fully described by four equations:
\begin{eqnarray}
0&=&\Lambda \beta^{(L)}_{{\ell}}+{i} m\left[(\ell-1)(\ell+2)\chi^{(L)}_{{\ell}}
+{\tilde\alpha}^{(L)}_{{\ell}}+\eta^{(L)}_{{\ell}}\right] ,\nn\\\label{axial_dec}\\
0&=&\Lambda t_{{\ell}}+{i} m g_{{\ell}} , 
\end{eqnarray}
where $L=0,1,2$. The choice $L=2$ refers to one of the Maxwell
equations, and
$\{\beta^{(L)}_{{\ell}},\,\chi^{(L)}_{{\ell}},\,{\tilde\alpha}^{(L)}_{{\ell}},\,\eta^{(L)}_{{\ell}},\,t_{{\ell}},\,g_{{\ell}}\}$
are combinations of the perturbation functions
$\{h^\ell_{0,1},\,u^\ell_{(4)}\}$, whose explicit form can be found
online~\cite{webpage}.  Actually, Einstein's equations imply Maxwell's
equations, and only three out of the four equations above are
independent. This can be easily verified as a consistency check of our
approach. The three independent equations can be solved for the
functions $h_0^\ell$, $h_1^\ell$ and $u_{(4)}^\ell$.  We define the
Regge-Wheeler function $\Psi^\ell$ as
\begin{equation}
\Psi^\ell=\frac{Fh_1^\ell}{r}\,.
% h_1^\ell=\frac{r\Psi^\ell}{F}\,.
\end{equation}
Then, from Eq.~\eqref{axial_dec} with $L=1$, we get
%%%
\begin{eqnarray}
&& {h_0^\ell}'=\frac{2 F(\Lambda r \omega  h_0^\ell-
2 Q \omega  u_{(4)}^\ell)+{i} \Lambda r \left(F \left(\Lambda
-2\right)-r^2 \omega ^2\right) \Psi^\ell }{F \Lambda r^2 \omega }\nn\\
&&-\frac{{i} m \tilde a M^2}{F \Lambda^2 M r^4 \omega ^2}
\left[12 {i} F \Lambda M \omega  h_0^\ell+4 {i} F^2 \Lambda Q 
\omega  u_{(4)}^\ell\right.\nn\\
&&\left.+(F-1) \Lambda^2 r \left(F \left(\Lambda-2\right)+r^2 
\omega ^2\right) \Psi^\ell +8 {i} F^2 Q r \omega  {u_{(4)}^\ell}'\right]\,.\nn
\end{eqnarray}
%%%%
Replacing this equation in the remaining Eqs.~\eqref{axial_dec} with
$L=0$ and $L=2$ we get a system for $u_{(4)}^\ell$ and $\Psi^\ell$
only. At first order in $\tilde{a}$, the system contains third
derivatives of $u_{(4)}^\ell$. Within our perturbative scheme, the
latter can be eliminated by using the zeroth order perturbation
equations. The result is a system of coupled, second-order equations
for $u_{(4)}^\ell$ and $\Psi^\ell$. In order to decouple the system at
zeroth order, we define a linear combination of two new functions
$Z_i^{(-)}$ ($i=1,2$), such that
%%%
\begin{eqnarray}
 u_4^\ell&=&\alpha_{11} Z_1^{(-)}+\alpha_{12} Z_2^{(-)},\\
 \Psi^\ell&=&\alpha_{21} Z_1^{(-)}+\alpha_{22} Z_2^{(-)},
\end{eqnarray}
%%%
where $\alpha_{21}={4i \alpha_{11}\omega Q}/({\Lambda q_1})$, $\alpha_{22}={4i
  \alpha_{12}\omega Q}/({\Lambda q_2})$, and we have defined
\begin{equation}
 q_{1,2}=3 M\pm\sqrt{9 M^2+4 (\Lambda-2) Q^2}, \label{q12}
\end{equation}
such that $q_1 q_2=-4 Q^2(\Lambda-2)$ and $q_1+q_2=6M$. The constants
$\alpha_{11}$ and $\alpha_{12}$ can be set equal to unity without loss
of generality.

Replacing the linear combinations above into the equations for
$u_{(4)}^\ell$ and $\Psi^\ell$ and solving for $Z_1^{(-)}$ and
$Z_2^{(-)}$, we obtain Eq.~\eqref{eF} with a ``minus''
superscript.

\begin{widetext}
%%%%%%%%%%%%%%
%\section{Potentials}\label{app:potentials}
%%%%%%%%%%%%%
The potentials appearing in the axial sector of the perturbation
equations~\eqref{eF} read
%%%%%%
\begin{eqnarray}
 V_0^{(i,-)}&=&\frac{F}{r^3}\left[-q_j+\frac{4Q^2}{r}\right],\\
%%%%%
 V_1^{(i,-)}&=& -\frac{M }{\Lambda (q_i-q_j) r^{11} \omega } \left[4 Q^6 \left(10 q_i q_j-12 M r+(-18 q_j+\Lambda (q_i+8 q_j)) r+\lambda r^2\right)\right.\nn\\
&&\left.-2 M q_j r^3 (r-2M) \left(r \left(\left(18+\Lambda\right) q_i+12 \left(\Lambda-3\right) r\right)-6 M (7 q_i+(-14+5 \Lambda) r)\right)\right.\nn\\
&&\left.+r^7 (2 M q_i q_j-4 \Lambda M (q_i-q_j) r-q_i q_j r) \omega ^2+Q^4 r \left(192 M^2 r-4 M \left(50 q_i q_j+(-92 q_j+\Lambda (4 q_i+39 q_j)) r+4 (6+\lambda) r^2\right)\right.\right.\nn\\
&&\left.\left.+r \left(\left(72+\Lambda\right) q_i q_j+8 \Lambda q_i r+4 r \left(2 (-16+7 \Lambda) q_j+\lambda r+\left(6+\Lambda\right) r^3 \omega ^2\right)\right)\right)\right.\nn\\
&&\left.+Q^2 r^2 \left(-192 M^3 r+4 M^2 \left(81 q_i q_j+(-154 q_j+\Lambda (4 q_i+61 q_j)) r+4 (12+\lambda) r^2\right)\right.\right.\nn\\
&&\left.\left.+r^2 \left(8 (3 \Lambda-7) q_j r+2 r^3 \left(-\Lambda q_j+2 \left(6+\Lambda\right) r\right) \omega ^2+q_i \left(q_j \left(32+\Lambda-r^2 \omega ^2\right)+2 \Lambda r \left(2+r^2 \omega ^2\right)\right)\right)\right.\right.\nn\\
&&\left.\left.-4 M r \left(\left(55+\Lambda\right) q_i q_j+4 \Lambda q_i r+r \left((-102+41 \Lambda) q_j+2 r \left(6+6 r^2 \omega ^2+\Lambda \left(-2+\Lambda+r^2 \omega ^2\right)\right)\right)\right)\right)\right],\\
%%%%%
 V_2^{(i,-)}&=& \frac{2 M F^2}{\Lambda (q_i-q_j) r^6 \omega }\left(Q^2 \left(q_i (5 q_j-4 r)+4 \left(6 M+(\Lambda-1) q_j\right) r\right)+q_j r (3 q_i r-2 M (4 q_i+3 \Lambda r))\right) ,\\
 W_1^{(i,-)}&=& \frac{4 \left(\Lambda-2\right) M F}{\Lambda (q_i-q_j) q_j^2 r^9 \omega } \left[4 Q^4 \left(10 q_j^2+9 \left(\Lambda-2\right) q_j r+r (\lambda r-12 M)\right)+q_j r^2 \left(2 M \left(r \left(-\left(18+\Lambda\right) q_j-12 \left(\Lambda-3\right) r\right)\right.\right.\right.\nn\\
&&\left.\left.\left.+6 M (7 q_j+(-14+5 \Lambda) r)\right)-q_j r^4 \omega ^2\right)+Q^2 r \left(96 M^2 r+q_j r \left(\left(32+\Lambda\right) q_j+28 \left(\Lambda-2\right) r\right)\right.\right.\nn\\
&&\left.\left.+4 M \left(-30 q_j^2+(56-25 \Lambda) q_j r-2 (6+\lambda) r^2\right)+4 \left(6+\Lambda\right) r^5 \omega ^2\right)\right],\\
 W_2^{(i,-)}&=&\frac{8 \left(\Lambda-2\right) M F^2}{\Lambda (q_i-q_j) q_j^2 r^6 \omega }\left(q_j r (8 M q_j+6 \Lambda M r-3 q_j r)+Q^2 \left(-5 q_j^2-4 \left(6 M+\left(\Lambda-2\right) q_j\right) r\right)\right) \,,
%%%%%%%%
\end{eqnarray}
where $q_{i,j}$ ($i,j=1,2$ and $i\neq j$) are defined as in
Eq.~\eqref{q12}, and $\lambda\equiv \Lambda (\ell+2)(\ell-1)$.
\end{widetext}

%%%%%%%%%%%%%%%%%%%%%%%%%%%%%%%%%%%%%%%%%%%%%%%%%%%%%%%
\subsection{Polar sector}\label{app:polar}
%%%%%%%%%%%%%%%%%%%%%%%%%%%%%%%%%%%%%%%%%%%%%%%%%%%%%%%

The equations governing the polar sector are more involved. They can
obtained by following Zerilli's original derivation of the
gravitational and electromagnetic perturbations of a RN
BH~\cite{Zerilli:1974ai}, extended to first order in the BH spin.

Polar perturbations are fully described by the equations:
%%%
\begin{eqnarray}
&0=&A^{(I)}_{{\ell}}+{i} mC^{(I)}_{{\ell}}, \label{dec_A}\\
&0=&\Lambda \alpha^{(L)}_{{\ell}}+{i} m\left[(\ell-1)(\ell+2)\xi^{(L)}_{{\ell}}
-{\tilde\beta}^{(L)}_{{\ell}}-\zeta^{(L)}_{{\ell}}\right],\nonumber\\
&& \label{dec_alpha}\\
&0=&\Lambda s_{{\ell}}-{i} m f_{{\ell}},\label{dec_s}
\end{eqnarray}
%%%%
where $I=0,...,5$ and $L=0,1,2$, and
$\{A^{(I)}_{{\ell}},\,C^{(I)}_{{\ell}},\,{\alpha}^{(L)}_{{\ell}},\,\xi^{(L)}_{{\ell}},\,{\tilde\beta}^{(L)}_{{\ell}},\,\zeta^{(L)}_{{\ell}},\,s_{{\ell}},\,f_{{\ell}}\}$
are combinations of the perturbation functions
$\{H^\ell_{0,1,2},\,K^\ell,\,u^\ell_{(1,2,3)}\}$, whose explicit form
can be found online~\cite{webpage}. Actually, only seven out of the
ten equations above are independent and they can be solved for the
seven polar functions: $H_0^\ell$, $H_1^\ell$, $H_2^\ell$, $K^\ell$,
$u_{(i)}^\ell$ ($i=1,2,3$).

The key point of Zerilli's calculation is to use perturbations of the
field strenght $F_{\mu\nu}$, rather than the electromagnetic potential
$A_\mu$, as dynamical variables. Following \cite{Zerilli:1974ai}, we
define
%%%
\begin{equation}
  \delta F_{\mu\nu}\equiv f_{\mu\nu}=\partial_\mu \delta A_\nu-\partial_\nu \delta A_\mu\,. \label{deff}
\end{equation}
%%%
In the polar sector, we fix the gauge by requiring
$u_{(3)}^\ell=0$; the remaining components are related to
$f_{\mu\nu}$ in the following way
\begin{eqnarray}
 u_{(1)}^\ell&=&r \tilde{f}_{02},\\
 u_{(2)}^\ell&=&F(r) \tilde{f}_{12},\\
 {u_{(1)}^\ell}'&=&r \tilde{f}_{01} + \tilde{f}_{02} - i r \omega \tilde{f}_{02}\,,
\end{eqnarray}
where $-\tilde{f}_{\mu\nu}$ denotes the angle-independent part of
$f_{\mu\nu}$ (note that our definition differs from Zerilli's
definition by a minus sign). The equation
%%%
\begin{equation}
 \tilde{f}_{01}=\tilde{f}_{02}'+i\omega \tilde{f}_{12},\label{eqftilde}
\end{equation}
%%%
is automatically satisfied due to Eq.~\eqref{deff}.

First, we solve Eq.~\eqref{dec_s} for $H_2^\ell$ and substitute the
result into the remaining equations. Then, Eq.~\eqref{dec_A} with
$I=5$ and Eq.~\eqref{dec_alpha} with $L=2$ can be solved for
$\tilde{f}_{01}$ and $\tilde{f}_{02}$. The solutions can be replaced
into Eq.~\eqref{eqftilde}, which takes the form of a second-order
differential equation for $\tilde{f}_{12}$ with source terms
linear in the polar gravitational function $H_0^\ell$,
$H_1^\ell$, $K^\ell$ and in their first derivatives.

The equations for the gravitational sector can be obtained by solving
Eq.~\eqref{dec_A} with $I=1$ and Eqs.~\eqref{dec_alpha} with $L=0,1$
for ${H_0^\ell}'$, ${H_1^\ell}'$ and ${K^\ell}'$. Similarly to the
axial sector, second derivatives of these perturbation functions
(which appear at first order in $\tilde{a}$) can be eliminated using
the zeroth-order equations. By substituing the solution into
Eq.~\eqref{dec_A} with $I=2$, one can solve for the function
$H_0^\ell$ and eliminate it from the remaining equations. As a result
of this procedure we obtain a system of coupled equations
%%%
\begin{eqnarray}
 \mathbf{Y}'+\mathbf{U}\mathbf{Y}=0,
\end{eqnarray}
%%%
where $\mathbf{Y}=(H_1^\ell,K^\ell,\tilde{f}_{12},\tilde{f}_{12}')$
and $\mathbf{U}$ is a matrix. The system above collectively denotes
two first-order equations for the gravitational perturbations
$H_0^\ell$ and $K^\ell$ and the second-order equation for
$\tilde{f}_{12}$, which has been separated into two first order
equations for $\tilde{f}_{12}$ and $\tilde{f}_{12}'$.
%%%%

As shown by Zerilli~\cite{Zerilli:1974ai}, at zeroth order in the BH
spin the two equations for the gravitational perturbations $H_0^\ell$
and $K^\ell$ can be reduced to a single, second-order equation. We
describe the procedure here, extending it to first order in the BH
spin.  The perturbations $H_0^\ell$ and $K^\ell$ satisfy the schematic
matrix-valued equation
\begin{equation}
 \mathbf{y}'=\mathbf{A}\mathbf{y}+\mathbf{S},\label{eqy}
\end{equation}
%%%
where $\mathbf{A}$ is a matrix, $\mathbf{y}=(K^\ell,H_1^\ell/\omega )$
and $\mathbf{S}=(S_1,S_2)$ is a source term, which depends on
$\tilde{f}_{12}$ and its derivatives only.  The idea is to find a
transformation $\mathbf{y}=\mathbf{F} \mathbf{\hat y}$ to a new pair
of functions $\mathbf{\hat y}=(\hat \Psi_1,\hat \Psi_2)$ such that
%%%
\begin{eqnarray}
 \frac{d \hat \Psi_1}{d\hat r}&=&G_1\hat \Psi_1+(1+W)\hat \Psi_2+\hat S_1, \label{eqHat1}\\
 \frac{d \hat \Psi_2}{d\hat r}&=&-(\omega^2-V-V_1)\hat \Psi_1+G_2\hat \Psi_2+\hat S_2 \,,
\label{eqHat2}
\end{eqnarray}
%%%%
where $\hat r$ is a new variable defined by $dr/d\hat r=n(r)$.
Indeed, assuming the relations above, we can solve Eq.~\eqref{eqHat1}
for $\hat\Psi_2$ and substitute it into Eq.~\eqref{eqHat2}. We obtain
a single, second-order equation for $\hat \Psi_1$:
%%%
\begin{equation}
 \frac{d^2\hat \Psi_1}{d{\hat r}^2}+\hat U\frac{d\hat \Psi_1}{d{\hat r}}
+\hat V\hat \Psi_1=\hat {\cal S}	,\label{eqR1}
\end{equation}
where, to first order,
%%%
\begin{eqnarray}
 \hat U&=&-\frac{G_1+G_2+F W'}{F'F},\\
 \hat V&=&\left(\omega ^2-V\right)(1+W)-V_1-F G_1',\\
 \hat {\cal S}&=&(1+W)\hat S_2+F\hat S_1'-\hat S_1 \left(G_2+F W'\right)\,.
\end{eqnarray}
%%%%
Note that $G_i$, $V_1$ and $W$ are first-order quantities in rotation,
which are absent in the nonrotating case discussed in Ref.~\cite{Zerilli:1974ai}.

Let us now find the explicit form of the transformation. Using
Eq.~\eqref{eqy}, we get
%%%
\begin{equation}
 \frac{d\mathbf{\hat y}}{d\hat r}=n(r)\mathbf{F}^{-1}\left[\mathbf{A}
\mathbf{F}-\frac{d\mathbf{F}}{dr}\right]+n(r)\mathbf{F}^{-1}\mathbf{S}\,.
\end{equation}
%%%
Therefore, the transformation matrix $\mathbf{F}$ and the remaining
functions must satisfy
\begin{equation}
 n(r)\mathbf{F}^{-1}\left[\mathbf{A}\mathbf{F}-\frac{d\mathbf{F}}{dr}\right]=\left(\begin{array}{cc}
                                                                                    G_1 & 1+W \\
                                                                                    -\omega^2+V+V_1 & G_2
                                                                                   \end{array}
\right), \label{matrixeq}
\end{equation}
%%%%
and $\mathbf{\hat S}=n \mathbf{F}^{-1}\mathbf{S}$. The equation above
can be solved perturbatively. At zeroth order it provides a system of
four equations that can be uniquely solved for the elements of the
matix $\mathbf{F}$, for $n(r)$ and for the potential $V$. We get
%%%%
\begin{eqnarray}
 \mathbf{F}_{11}&=& \left[{2 r^3 \left(r \left(6 M+\left(\Lambda-2\right) r\right)-4 Q^2\right)}\right]^{-1}\nn\\
 &&\times\left[16 Q^4-4 Q^2 r \left(11 M+\left(\Lambda-4\right) r\right)\right.\nn\\
 &&\left.+r^2 \left(24 M^2+6 \left(\Lambda-2\right) M r+\lambda r^2\right)\right]  ,\nn\\
 \mathbf{F}_{12}&=& 1    ,\nn\\\
 \mathbf{F}_{21}&=& -i \left[1+\frac{Q^2-M r}{r^2F}+\frac{8 Q^2-6
 M r}{r \left[6 M+\left(\Lambda-2\right) r\right]-4 Q^2}\right]   ,\nn\\\
 \mathbf{F}_{22}&=&  -\frac{i r}{F}  ,\nn\
 \end{eqnarray}
 and
 \begin{eqnarray}
 V(r)&=& \frac{F}{r^2} \left(r \left(6 M+\left(\Lambda-2\right) r\right)-4 Q^2\right)^{-2}\nn\\
 &&\times \left[-32 Q^6+24 Q^4 r \left(6 M+\left(\Lambda-2\right) r\right)\right.\nn\\
 &&\left.+8 Q^2 r^2 \left(3 \left(\Lambda-2\right) r^2-27 M^2-2 (4\Lambda-11) M r\right)\right.\nn\\
 &&\left.+r^3 \left(72 M^3+36 \left(\Lambda-2\right) M^2 r+6 \left(\Lambda-2\right)^2 M r^2\right.\right.\nn\\
 &&\left.\left.+\Lambda \left(\Lambda-2\right)^2 r^3\right)\right],\\
  n(r)&=&F(r).\label{abovelambda}
\end{eqnarray}
%%%
Note that, by virtue of the field equations, the coordinate $\hat r$
is the standard tortoise coordinate. The equations above correct some
typos in Ref.~\cite{Zerilli:1974ai}. At first order, we can solve the
four equations~\eqref{matrixeq} for the variables $G_1$, $G_2$, $W$
and $V_1$. The lengthy form of these solutions is presented
online~\cite{webpage}.
%%%

With the transformation at hand, we can compute the explicit form of
Eq.~\eqref{eqR1}. The source term $\hat {\cal S}$ explicitly
depends on the second derivatives of $\tilde f_{12}$ and, upon
substitution of the perturbation equation for $\tilde{f}_{12}$, it
depends on $K^\ell$ and $H_1^\ell$. This dependence can be eliminated
using the definitions
%%%
\begin{eqnarray}
 K^\ell&=& \mathbf{F}_{11}\hat\Psi_1+\frac{\mathbf{F}_{12}}{1+W}\left[F\hat\Psi_1'-G_1 \hat\Psi_1-\hat S_1\right]\,, \nn \\
 H_1^\ell&=& \omega\mathbf{F}_{21}\hat\Psi_1+\frac{\omega\mathbf{F}_{22}}{1+W}\left[F\hat\Psi_1'-G_1 \hat\Psi_1-\hat S_1\right], \nn
\end{eqnarray}
%%%
which allows us to write $K^\ell$ and $H_1^\ell$ in terms of
$\hat\Psi_1$ and its first derivative only. When inserted into
Eq.~\eqref{eqR1}, the equations above introduce extra coefficients in
front of ${d\hat \Psi_1}/{dr_*}$ and $\hat \Psi_1$. The final result
is an equation of the same form as Eq.~\eqref{eqR1}, with coefficients
given in~\cite{webpage}.

To summarize, we have obtained two second-order equations for the
functions $\hat \Psi_1$ and $\tilde{f}_{12}$ which describe
gravitational and electromagnetic perturbations, respectively. At
zeroth order in rotation, these equations can be
decoupled~\cite{Moncrief:1975sb,Chandra} by introducing the
%first-order corrections couple the gravitational and electromagnetic sectors
%together. In order to decouple the system at zeroth order, we define a linear
%combination of the new 
functions $Z_i^{(+)}$ such that
%%%
\begin{eqnarray}
 \hat\Psi_1&=& B_{11} Z_1^{(-)}+B_{12} Z_2^{(+)},\\
 \tilde{f}_{EM}&=&B_{21} Z_1^{(+)}+B_{22} Z_2^{(+)},
\end{eqnarray}
%%%
where $B_{ij}$ are functions of $r$ and we defined $\tilde{f}_{EM}\equiv
F\tilde{f}_{12}$. It is straightforward to verify that with the choice
%%%
\begin{eqnarray}
\frac{B_{11}}{\alpha}+q_2&=&\frac{B_{12}}{\beta}+q_1=\frac{4 Q^2}{r} ,\\
% \frac{B_{12}}{\beta}&=& -q_1+\frac{4 Q^2}{r} ,\\
\frac{B_{21}}{\alpha}&=&\frac{B_{22}}{\beta}= -\frac{8 i Q}{\omega },
% B_{22}&=& -\frac{8 i Q \beta }{\omega },\\
\end{eqnarray}
%%%%
the final set of equations takes the form of Eq.~\eqref{eF} with a
``plus'' superscript, i.e., the equations are decoupled at zeroth
order (but coupled at first order) in $\tilde a$.  In the equations
above, $\alpha$ and $\beta$ are constants that can be set to unity
without loss of generality.
Our equations reduce to those obtained by Chandrasekhar~\cite{Chandra}
in the nonspinning case, as they should. Let us stress again that the
derivation sketched in this Appendix and in \cite{webpage} corrects
some typos in Ref.~\cite{Zerilli:1974ai}.
%%%%%%%%%%%%%%%%%%%%%%%%%%%

The potentials appearing in the perturbation equations~\eqref{eF} of
the polar sector are very lengthy~\cite{webpage}, and their practical
use may be limited. However, since - according to our numerical
evidence - axial and polar modes are isospectral to linear order in
$\tilde a$, the potentials describing axial
perturbations (explicitly listed in Appendix \ref{app:axial}) are
sufficient to compute the entire QNM spectrum of slowly rotating KN
BHs.

%%%%%%%%%%%%%%
\section{Coefficients of the recurrence relation for scalar QNMs of a KN BHs}\label{app:CFs}
%%%%%%%%%%%%%
\begin{widetext}
Using the same ansatz as in Eq.~\eqref{ansatzCF}, Eq.~\eqref{scalar} reduce to a five-term recurrence relation
%%%
\begin{eqnarray}
&&\alpha_0 a_{1} + \beta_0 a_{0} = 0\,,  \quad  \hspace{6.3cm} n =0\,,\nn\\
%%%
&&\alpha_1 a_{2} + \beta_1 a_{1} + \gamma_1 a_{0} = 0\,, \quad  \hspace{5.2cm} n =1\,,\nn \\
%%%
&&\alpha_2 a_{3} + \beta_2 a_{2} + \gamma_2 a_{1} + \delta_2 a_{0} = 0\,, \quad  \hspace{4.15cm} n =2\,,\nn \\
%%%
&&\alpha_n a_{n+1} + \beta_n a_{n} + \gamma_n a_{n-1} + \delta_n a_{n-2} + \rho_n a_{n-3} = 0\,, \quad \hspace{1.28cm} n > 2\,,\nn 
\end{eqnarray}
%%%%
whose coefficients to first order in $\tilde{a}$, using Leaver's $2M=1$
unit convention, read:
\begin{eqnarray}
 \alpha_n&=& (1+n) \Delta_Q (1+\Delta_Q)^2 \left(2 (1+n) \Delta_Q-i (1+\Delta_Q)^2 \omega \right)+2 i m \tilde{a}(1+n) \Delta_Q (1+\Delta_Q)^2 \,,\\
 %%%
 \beta_n&=& \Delta_Q (1+\Delta_Q) \left(-2 \Delta_Q \left(-1+4 n^2+3 \Delta_Q+\ell  (1+\ell ) (1+\Delta_Q)+n (-2+6 \Delta_Q)\right)\right.\nn\\
 &&\left.+i (1+\Delta_Q)^2 (-1+5 \Delta_Q+4 n (1+\Delta_Q)) \omega +2 (1+\Delta_Q)^4 \omega ^2\right)\nn\\
 &&-2 m\tilde{a} \Delta_Q (1+\Delta_Q) (i (4 n+3 \Delta_Q-1)+(1+\Delta_Q) (2+\Delta_Q) \omega )\,,\\
 %%%
 \gamma_n&=& \Delta_Q \left(2 \Delta_Q \left(6 n (\Delta_Q-1) (3+\Delta_Q)-2 n^2 \left(\Delta_Q^2-3\right)-(\Delta_Q-1) (11-7 \Delta_Q+2 \ell  (1+\ell ) (1+\Delta_Q))\right)\right.\nn\\
 &&\left.+i (1+\Delta_Q)^2 \left(9+6 \Delta_Q-15 \Delta_Q^2+2 n (-3+\Delta_Q (-6+5 \Delta_Q))\right) \omega +2 (1+\Delta_Q)^4 (-3+2 \Delta_Q) \omega ^2\right)\nn\\
 &&+2 m  \tilde{a}\Delta_Q \left(3 i (\Delta_Q-1) (3+\Delta_Q)-2 i n \left(\Delta_Q^2-3\right)-2 (1+\Delta_Q) \left(\Delta_Q^2-3\right) \omega \right)\,,\\
 %%%
 \delta_n&=& (\Delta_Q-1) \Delta_Q \left(2 \Delta_Q \left(29+4 n^2+\ell +\ell ^2-\left(21+\ell +\ell ^2\right) \Delta_Q+n (-22+6 \Delta_Q)\right)\right.\nn\\
 &&\left.+i (1+\Delta_Q) (11+(38-17 \Delta_Q) \Delta_Q+4 n (-1+(-4+\Delta_Q) \Delta_Q)) \omega +2 (\Delta_Q-3) (1+\Delta_Q)^3 \omega ^2\right)\nn\\
 &&+2 m \tilde{a} (\Delta_Q-1) \Delta_Q (-11 i+4 i n+6 \omega +\Delta_Q (3 i-(\Delta_Q-3) \omega )) \,,\\
 \rho_n&=& (\Delta_Q-1)^2 \Delta_Q (-4+n-2 i \omega ) \left(2 (-4+n) \Delta_Q-i (1+\Delta_Q)^2 \omega \right)+2 i m\tilde{a} (\Delta_Q-1)^2 \Delta_Q (-4+n-2 i \omega )\,,
\end{eqnarray}
%%%
where $\Delta_Q=\sqrt{1-4Q^2}$.
\end{widetext}
%%

%%%%%
%\bibliographystyle{myutphys} %Needed by Leonardo
\bibliography{KN}

%merlin.mbs apsrev4-1.bst 2010-07-25 4.21a (PWD, AO, DPC) hacked
%Control: key (0)
%Control: author (8) initials jnrlst
%Control: editor formatted (1) identically to author
%Control: production of article title (-1) disabled
%Control: page (0) single
%Control: year (1) truncated
%Control: production of eprint (0) enabled
\begin{thebibliography}{82}%
\makeatletter
\providecommand \@ifxundefined [1]{%
 \@ifx{#1\undefined}
}%
\providecommand \@ifnum [1]{%
 \ifnum #1\expandafter \@firstoftwo
 \else \expandafter \@secondoftwo
 \fi
}%
\providecommand \@ifx [1]{%
 \ifx #1\expandafter \@firstoftwo
 \else \expandafter \@secondoftwo
 \fi
}%
\providecommand \natexlab [1]{#1}%
\providecommand \enquote  [1]{``#1''}%
\providecommand \bibnamefont  [1]{#1}%
\providecommand \bibfnamefont [1]{#1}%
\providecommand \citenamefont [1]{#1}%
\providecommand \href@noop [0]{\@secondoftwo}%
\providecommand \href [0]{\begingroup \@sanitize@url \@href}%
\providecommand \@href[1]{\@@startlink{#1}\@@href}%
\providecommand \@@href[1]{\endgroup#1\@@endlink}%
\providecommand \@sanitize@url [0]{\catcode `\\12\catcode `\$12\catcode
  `\&12\catcode `\#12\catcode `\^12\catcode `\_12\catcode `\%12\relax}%
\providecommand \@@startlink[1]{}%
\providecommand \@@endlink[0]{}%
\providecommand \url  [0]{\begingroup\@sanitize@url \@url }%
\providecommand \@url [1]{\endgroup\@href {#1}{\urlprefix }}%
\providecommand \urlprefix  [0]{URL }%
\providecommand \Eprint [0]{\href }%
\providecommand \doibase [0]{http://dx.doi.org/}%
\providecommand \selectlanguage [0]{\@gobble}%
\providecommand \bibinfo  [0]{\@secondoftwo}%
\providecommand \bibfield  [0]{\@secondoftwo}%
\providecommand \translation [1]{[#1]}%
\providecommand \BibitemOpen [0]{}%
\providecommand \bibitemStop [0]{}%
\providecommand \bibitemNoStop [0]{.\EOS\space}%
\providecommand \EOS [0]{\spacefactor3000\relax}%
\providecommand \BibitemShut  [1]{\csname bibitem#1\endcsname}%
\let\auto@bib@innerbib\@empty
%</preamble>
\bibitem [{\citenamefont {Pani}\ \emph {et~al.}(2013)\citenamefont {Pani},
  \citenamefont {Berti},\ and\ \citenamefont {Gualtieri}}]{Pani:2013ija}%
  \BibitemOpen
  \bibfield  {author} {\bibinfo {author} {\bibfnamefont {P.}~\bibnamefont
  {Pani}}, \bibinfo {author} {\bibfnamefont {E.}~\bibnamefont {Berti}}, \ and\
  \bibinfo {author} {\bibfnamefont {L.}~\bibnamefont {Gualtieri}},\ }\href
  {\doibase 10.1103/PhysRevLett.110.241103} {\bibfield  {journal} {\bibinfo
  {journal} {Phys.Rev.Lett.}\ }\textbf {\bibinfo {volume} {110}},\ \bibinfo
  {pages} {241103} (\bibinfo {year} {2013})},\ \Eprint
  {http://arxiv.org/abs/1304.1160} {arXiv:1304.1160 [gr-qc]} \BibitemShut
  {NoStop}%
%%CITATION = ARXIV:1304.1160;%%
\bibitem [{\citenamefont {Chrusciel}\ \emph {et~al.}(2012)\citenamefont
  {Chrusciel}, \citenamefont {Costa},\ and\ \citenamefont
  {Heusler}}]{Chrusciel:2012jk}%
  \BibitemOpen
  \bibfield  {author} {\bibinfo {author} {\bibfnamefont {P.~T.}\ \bibnamefont
  {Chrusciel}}, \bibinfo {author} {\bibfnamefont {J.~L.}\ \bibnamefont
  {Costa}}, \ and\ \bibinfo {author} {\bibfnamefont {M.}~\bibnamefont
  {Heusler}},\ }\href@noop {} {\bibfield  {journal} {\bibinfo  {journal}
  {Living Rev.Rel.}\ }\textbf {\bibinfo {volume} {15}},\ \bibinfo {pages} {7}
  (\bibinfo {year} {2012})},\ \Eprint {http://arxiv.org/abs/1205.6112}
  {arXiv:1205.6112 [gr-qc]} \BibitemShut {NoStop}%
%%CITATION = ARXIV:1205.6112;%%
\bibitem [{\citenamefont {Newman}\ \emph {et~al.}(1965)\citenamefont {Newman},
  \citenamefont {Couch}, \citenamefont {Chinnapared}, \citenamefont {Exton},
  \citenamefont {Prakash} \emph {et~al.}}]{Newman:1965my}%
  \BibitemOpen
  \bibfield  {author} {\bibinfo {author} {\bibfnamefont {E.~T.}\ \bibnamefont
  {Newman}}, \bibinfo {author} {\bibfnamefont {R.}~\bibnamefont {Couch}},
  \bibinfo {author} {\bibfnamefont {K.}~\bibnamefont {Chinnapared}}, \bibinfo
  {author} {\bibfnamefont {A.}~\bibnamefont {Exton}}, \bibinfo {author}
  {\bibfnamefont {A.}~\bibnamefont {Prakash}},  \emph {et~al.},\ }\href@noop {}
  {\bibfield  {journal} {\bibinfo  {journal} {J.Math.Phys.}\ }\textbf {\bibinfo
  {volume} {6}},\ \bibinfo {pages} {918} (\bibinfo {year} {1965})}\BibitemShut
  {NoStop}%
%%CITATION = JMAPA,6,918;%%
\bibitem [{\citenamefont {Carter}(1968)}]{Carter:1968rr}%
  \BibitemOpen
  \bibfield  {author} {\bibinfo {author} {\bibfnamefont {B.}~\bibnamefont
  {Carter}},\ }\href {\doibase 10.1103/PhysRev.174.1559} {\bibfield  {journal}
  {\bibinfo  {journal} {Phys.Rev.}\ }\textbf {\bibinfo {volume} {174}},\
  \bibinfo {pages} {1559} (\bibinfo {year} {1968})}\BibitemShut {NoStop}%
%%CITATION = PHRVA,174,1559;%%
\bibitem [{\citenamefont {Pekeris}\ and\ \citenamefont
  {Frankowski}(1989)}]{Pekeris:1989mu}%
  \BibitemOpen
  \bibfield  {author} {\bibinfo {author} {\bibfnamefont {C.}~\bibnamefont
  {Pekeris}}\ and\ \bibinfo {author} {\bibfnamefont {K.}~\bibnamefont
  {Frankowski}},\ }\href {\doibase 10.1103/PhysRevA.39.518} {\bibfield
  {journal} {\bibinfo  {journal} {Phys.Rev.}\ }\textbf {\bibinfo {volume}
  {A39}},\ \bibinfo {pages} {518} (\bibinfo {year} {1989})}\BibitemShut
  {NoStop}%
%%CITATION = PHRVA,A39,518;%%
\bibitem [{\citenamefont {Gibbons}(1975)}]{Gibbons:1975kk}%
  \BibitemOpen
  \bibfield  {author} {\bibinfo {author} {\bibfnamefont {G.}~\bibnamefont
  {Gibbons}},\ }\href {\doibase 10.1007/BF01609829} {\bibfield  {journal}
  {\bibinfo  {journal} {Commun.Math.Phys.}\ }\textbf {\bibinfo {volume} {44}},\
  \bibinfo {pages} {245} (\bibinfo {year} {1975})}\BibitemShut {NoStop}%
%%CITATION = CMPHA,44,245;%%
\bibitem [{\citenamefont {Blandford}\ and\ \citenamefont
  {Znajek}(1977)}]{Blandford:1977ds}%
  \BibitemOpen
  \bibfield  {author} {\bibinfo {author} {\bibfnamefont {R.}~\bibnamefont
  {Blandford}}\ and\ \bibinfo {author} {\bibfnamefont {R.}~\bibnamefont
  {Znajek}},\ }\href@noop {} {\bibfield  {journal} {\bibinfo  {journal}
  {Mon.Not.Roy.Astron.Soc.}\ }\textbf {\bibinfo {volume} {179}},\ \bibinfo
  {pages} {433} (\bibinfo {year} {1977})}\BibitemShut {NoStop}%
%%CITATION = MNRAA,179,433;%%
\bibitem [{\citenamefont {Dadhich}\ and\ \citenamefont
  {Turakulov}(2002)}]{Dadhich:2001sz}%
  \BibitemOpen
  \bibfield  {author} {\bibinfo {author} {\bibfnamefont {N.}~\bibnamefont
  {Dadhich}}\ and\ \bibinfo {author} {\bibfnamefont {Z.~Y.}\ \bibnamefont
  {Turakulov}},\ }\href@noop {} {\bibfield  {journal} {\bibinfo  {journal}
  {Class.Quant.Grav.}\ }\textbf {\bibinfo {volume} {19}},\ \bibinfo {pages}
  {2765} (\bibinfo {year} {2002})},\ \Eprint
  {http://arxiv.org/abs/gr-qc/0112031} {arXiv:gr-qc/0112031 [gr-qc]}
  \BibitemShut {NoStop}%
%%CITATION = GR-QC/0112031;%%
\bibitem [{\citenamefont {Unruh}(1973)}]{Unruh:1973}%
  \BibitemOpen
  \bibfield  {author} {\bibinfo {author} {\bibfnamefont {W.}~\bibnamefont
  {Unruh}},\ }\href {\doibase 10.1103/PhysRevLett.31.1265} {\bibfield
  {journal} {\bibinfo  {journal} {Phys. Rev. Lett.}\ }\textbf {\bibinfo
  {volume} {31}},\ \bibinfo {pages} {1265} (\bibinfo {year}
  {1973})}\BibitemShut {NoStop}%
\bibitem [{\citenamefont {Chandrasekhar}(1976)}]{Chandrasekhar:1976ap}%
  \BibitemOpen
  \bibfield  {author} {\bibinfo {author} {\bibfnamefont {S.}~\bibnamefont
  {Chandrasekhar}},\ }\href@noop {} {\bibfield  {journal} {\bibinfo  {journal}
  {Proc.Roy.Soc.Lond.}\ }\textbf {\bibinfo {volume} {A349}},\ \bibinfo {pages}
  {571} (\bibinfo {year} {1976})}\BibitemShut {NoStop}%
%%CITATION = PRSLA,A349,571;%%
\bibitem [{\citenamefont {Page}(1976)}]{Page:1976jj}%
  \BibitemOpen
  \bibfield  {author} {\bibinfo {author} {\bibfnamefont {D.~N.}\ \bibnamefont
  {Page}},\ }\href {\doibase 10.1103/PhysRevD.14.1509} {\bibfield  {journal}
  {\bibinfo  {journal} {Phys.Rev.}\ }\textbf {\bibinfo {volume} {D14}},\
  \bibinfo {pages} {1509} (\bibinfo {year} {1976})}\BibitemShut {NoStop}%
%%CITATION = PHRVA,D14,1509;%%
\bibitem [{\citenamefont {Torres~del Castillo}\ and\ \citenamefont
  {Silva-Ortigoza}(1990)}]{TorresdelCastillo:1990aw}%
  \BibitemOpen
  \bibfield  {author} {\bibinfo {author} {\bibfnamefont {G.}~\bibnamefont
  {Torres~del Castillo}}\ and\ \bibinfo {author} {\bibfnamefont
  {G.}~\bibnamefont {Silva-Ortigoza}},\ }\href {\doibase
  10.1103/PhysRevD.42.4082} {\bibfield  {journal} {\bibinfo  {journal}
  {Phys.Rev.}\ }\textbf {\bibinfo {volume} {D42}},\ \bibinfo {pages} {4082}
  (\bibinfo {year} {1990})}\BibitemShut {NoStop}%
%%CITATION = PHRVA,D42,4082;%%
\bibitem [{\citenamefont {Carter}\ and\ \citenamefont
  {Mclenaghan}(1979)}]{Carter:1979fe}%
  \BibitemOpen
  \bibfield  {author} {\bibinfo {author} {\bibfnamefont {B.}~\bibnamefont
  {Carter}}\ and\ \bibinfo {author} {\bibfnamefont {R.}~\bibnamefont
  {Mclenaghan}},\ }\href {\doibase 10.1103/PhysRevD.19.1093} {\bibfield
  {journal} {\bibinfo  {journal} {Phys.Rev.}\ }\textbf {\bibinfo {volume}
  {D19}},\ \bibinfo {pages} {1093} (\bibinfo {year} {1979})}\BibitemShut
  {NoStop}%
%%CITATION = PHRVA,D19,1093;%%
\bibitem [{\citenamefont {Carter}(2006)}]{Carter:2006hj}%
  \BibitemOpen
  \bibfield  {author} {\bibinfo {author} {\bibfnamefont {B.}~\bibnamefont
  {Carter}},\ }\href {\doibase 10.1063/1.2218167} {\bibfield  {journal}
  {\bibinfo  {journal} {AIP Conf.Proc.}\ }\textbf {\bibinfo {volume} {841}},\
  \bibinfo {pages} {29} (\bibinfo {year} {2006})},\ \Eprint
  {http://arxiv.org/abs/gr-qc/0604064} {arXiv:gr-qc/0604064 [gr-qc]}
  \BibitemShut {NoStop}%
%%CITATION = GR-QC/0604064;%%
\bibitem [{\citenamefont {Kokkotas}\ and\ \citenamefont
  {Schmidt}(1999)}]{Kokkotas:1999bd}%
  \BibitemOpen
  \bibfield  {author} {\bibinfo {author} {\bibfnamefont {K.~D.}\ \bibnamefont
  {Kokkotas}}\ and\ \bibinfo {author} {\bibfnamefont {B.~G.}\ \bibnamefont
  {Schmidt}},\ }\href@noop {} {\bibfield  {journal} {\bibinfo  {journal}
  {Living Rev.Rel.}\ }\textbf {\bibinfo {volume} {2}},\ \bibinfo {pages} {2}
  (\bibinfo {year} {1999})},\ \Eprint {http://arxiv.org/abs/gr-qc/9909058}
  {arXiv:gr-qc/9909058 [gr-qc]} \BibitemShut {NoStop}%
\bibitem [{\citenamefont {Nollert}(1999)}]{Nollert:1999ji}%
  \BibitemOpen
  \bibfield  {author} {\bibinfo {author} {\bibfnamefont {H.-P.}\ \bibnamefont
  {Nollert}},\ }\href {\doibase 10.1088/0264-9381/16/12/201} {\bibfield
  {journal} {\bibinfo  {journal} {Class.Quant.Grav.}\ }\textbf {\bibinfo
  {volume} {16}},\ \bibinfo {pages} {R159} (\bibinfo {year}
  {1999})}\BibitemShut {NoStop}%
%%CITATION = CQGRD,16,R159;%%
\bibitem [{\citenamefont {Ferrari}\ and\ \citenamefont
  {Gualtieri}(2008)}]{Ferrari:2007dd}%
  \BibitemOpen
  \bibfield  {author} {\bibinfo {author} {\bibfnamefont {V.}~\bibnamefont
  {Ferrari}}\ and\ \bibinfo {author} {\bibfnamefont {L.}~\bibnamefont
  {Gualtieri}},\ }\href {\doibase 10.1007/s10714-007-0585-1} {\bibfield
  {journal} {\bibinfo  {journal} {Gen.Rel.Grav.}\ }\textbf {\bibinfo {volume}
  {40}},\ \bibinfo {pages} {945} (\bibinfo {year} {2008})},\ \Eprint
  {http://arxiv.org/abs/0709.0657} {arXiv:0709.0657 [gr-qc]} \BibitemShut
  {NoStop}%
\bibitem [{\citenamefont {Berti}\ \emph {et~al.}(2009)\citenamefont {Berti},
  \citenamefont {Cardoso},\ and\ \citenamefont {Starinets}}]{Berti:2009kk}%
  \BibitemOpen
  \bibfield  {author} {\bibinfo {author} {\bibfnamefont {E.}~\bibnamefont
  {Berti}}, \bibinfo {author} {\bibfnamefont {V.}~\bibnamefont {Cardoso}}, \
  and\ \bibinfo {author} {\bibfnamefont {A.~O.}\ \bibnamefont {Starinets}},\
  }\href {\doibase 10.1088/0264-9381/26/16/163001} {\bibfield  {journal}
  {\bibinfo  {journal} {Class.Quant.Grav.}\ }\textbf {\bibinfo {volume} {26}},\
  \bibinfo {pages} {163001} (\bibinfo {year} {2009})},\ \Eprint
  {http://arxiv.org/abs/0905.2975} {arXiv:0905.2975 [gr-qc]} \BibitemShut
  {NoStop}%
%%CITATION = ARXIV:0905.2975;%%
\bibitem [{\citenamefont {Konoplya}\ and\ \citenamefont
  {Zhidenko}(2011)}]{Konoplya:2011qq}%
  \BibitemOpen
  \bibfield  {author} {\bibinfo {author} {\bibfnamefont {R.}~\bibnamefont
  {Konoplya}}\ and\ \bibinfo {author} {\bibfnamefont {A.}~\bibnamefont
  {Zhidenko}},\ }\href {\doibase 10.1103/RevModPhys.83.793} {\bibfield
  {journal} {\bibinfo  {journal} {Rev.Mod.Phys.}\ }\textbf {\bibinfo {volume}
  {83}},\ \bibinfo {pages} {793} (\bibinfo {year} {2011})},\ \Eprint
  {http://arxiv.org/abs/1102.4014} {arXiv:1102.4014 [gr-qc]} \BibitemShut
  {NoStop}%
%%CITATION = ARXIV:1102.4014;%%
\bibitem [{\citenamefont {Berti}\ and\ \citenamefont
  {Kokkotas}(2005)}]{Berti:2005eb}%
  \BibitemOpen
  \bibfield  {author} {\bibinfo {author} {\bibfnamefont {E.}~\bibnamefont
  {Berti}}\ and\ \bibinfo {author} {\bibfnamefont {K.~D.}\ \bibnamefont
  {Kokkotas}},\ }\href {\doibase 10.1103/PhysRevD.71.124008} {\bibfield
  {journal} {\bibinfo  {journal} {Phys.Rev.}\ }\textbf {\bibinfo {volume}
  {D71}},\ \bibinfo {pages} {124008} (\bibinfo {year} {2005})},\ \Eprint
  {http://arxiv.org/abs/gr-qc/0502065} {arXiv:gr-qc/0502065 [gr-qc]}
  \BibitemShut {NoStop}%
%%CITATION = GR-QC/0502065;%%
\bibitem [{\citenamefont {Konoplya}\ and\ \citenamefont
  {Zhidenko}(2013)}]{Konoplya:2013rxa}%
  \BibitemOpen
  \bibfield  {author} {\bibinfo {author} {\bibfnamefont {R.}~\bibnamefont
  {Konoplya}}\ and\ \bibinfo {author} {\bibfnamefont {A.}~\bibnamefont
  {Zhidenko}},\ }\href@noop {} {\  (\bibinfo {year} {2013})},\ \Eprint
  {http://arxiv.org/abs/1307.1812} {arXiv:1307.1812 [gr-qc]} \BibitemShut
  {NoStop}%
%%CITATION = ARXIV:1307.1812;%%
\bibitem [{\citenamefont {Jing}\ and\ \citenamefont {Pan}(2005)}]{Jing:2005pk}%
  \BibitemOpen
  \bibfield  {author} {\bibinfo {author} {\bibfnamefont {J.-l.}\ \bibnamefont
  {Jing}}\ and\ \bibinfo {author} {\bibfnamefont {Q.-y.}\ \bibnamefont {Pan}},\
  }\href {\doibase 10.1016/j.nuclphysb.2005.08.038} {\bibfield  {journal}
  {\bibinfo  {journal} {Nucl.Phys.}\ }\textbf {\bibinfo {volume} {B728}},\
  \bibinfo {pages} {109} (\bibinfo {year} {2005})},\ \Eprint
  {http://arxiv.org/abs/gr-qc/0506098} {arXiv:gr-qc/0506098 [gr-qc]}
  \BibitemShut {NoStop}%
%%CITATION = GR-QC/0506098;%%
\bibitem [{\citenamefont {Furuhashi}\ and\ \citenamefont
  {Nambu}(2004)}]{Furuhashi:2004jk}%
  \BibitemOpen
  \bibfield  {author} {\bibinfo {author} {\bibfnamefont {H.}~\bibnamefont
  {Furuhashi}}\ and\ \bibinfo {author} {\bibfnamefont {Y.}~\bibnamefont
  {Nambu}},\ }\href {\doibase 10.1143/PTP.112.983} {\bibfield  {journal}
  {\bibinfo  {journal} {Prog.Theor.Phys.}\ }\textbf {\bibinfo {volume} {112}},\
  \bibinfo {pages} {983} (\bibinfo {year} {2004})},\ \Eprint
  {http://arxiv.org/abs/gr-qc/0402037} {arXiv:gr-qc/0402037 [gr-qc]}
  \BibitemShut {NoStop}%
%%CITATION = GR-QC/0402037;%%
\bibitem [{\citenamefont {Hartman}\ \emph {et~al.}(2010)\citenamefont
  {Hartman}, \citenamefont {Song},\ and\ \citenamefont
  {Strominger}}]{Hartman:2009nz}%
  \BibitemOpen
  \bibfield  {author} {\bibinfo {author} {\bibfnamefont {T.}~\bibnamefont
  {Hartman}}, \bibinfo {author} {\bibfnamefont {W.}~\bibnamefont {Song}}, \
  and\ \bibinfo {author} {\bibfnamefont {A.}~\bibnamefont {Strominger}},\
  }\href {\doibase 10.1007/JHEP03(2010)118} {\bibfield  {journal} {\bibinfo
  {journal} {JHEP}\ }\textbf {\bibinfo {volume} {1003}},\ \bibinfo {pages}
  {118} (\bibinfo {year} {2010})},\ \Eprint {http://arxiv.org/abs/0908.3909}
  {arXiv:0908.3909 [hep-th]} \BibitemShut {NoStop}%
%%CITATION = ARXIV:0908.3909;%%
\bibitem [{\citenamefont {Hartman}\ \emph {et~al.}(2009)\citenamefont
  {Hartman}, \citenamefont {Murata}, \citenamefont {Nishioka},\ and\
  \citenamefont {Strominger}}]{Hartman:2008pb}%
  \BibitemOpen
  \bibfield  {author} {\bibinfo {author} {\bibfnamefont {T.}~\bibnamefont
  {Hartman}}, \bibinfo {author} {\bibfnamefont {K.}~\bibnamefont {Murata}},
  \bibinfo {author} {\bibfnamefont {T.}~\bibnamefont {Nishioka}}, \ and\
  \bibinfo {author} {\bibfnamefont {A.}~\bibnamefont {Strominger}},\ }\href
  {\doibase 10.1088/1126-6708/2009/04/019} {\bibfield  {journal} {\bibinfo
  {journal} {JHEP}\ }\textbf {\bibinfo {volume} {0904}},\ \bibinfo {pages}
  {019} (\bibinfo {year} {2009})},\ \Eprint {http://arxiv.org/abs/0811.4393}
  {arXiv:0811.4393 [hep-th]} \BibitemShut {NoStop}%
%%CITATION = ARXIV:0811.4393;%%
\bibitem [{\citenamefont {Teukolsky}\ and\ \citenamefont
  {Press}(1974)}]{Teukolsky:1974yv}%
  \BibitemOpen
  \bibfield  {author} {\bibinfo {author} {\bibfnamefont {S.}~\bibnamefont
  {Teukolsky}}\ and\ \bibinfo {author} {\bibfnamefont {W.}~\bibnamefont
  {Press}},\ }\href {\doibase 10.1086/153180} {\bibfield  {journal} {\bibinfo
  {journal} {Astrophys.J.}\ }\textbf {\bibinfo {volume} {193}},\ \bibinfo
  {pages} {443} (\bibinfo {year} {1974})}\BibitemShut {NoStop}%
%%CITATION = ASJOA,193,443;%%
\bibitem [{\citenamefont {Wagh}\ and\ \citenamefont
  {Dadhich}(1985)}]{Wagh:1986cz}%
  \BibitemOpen
  \bibfield  {author} {\bibinfo {author} {\bibfnamefont {S.}~\bibnamefont
  {Wagh}}\ and\ \bibinfo {author} {\bibfnamefont {N.}~\bibnamefont {Dadhich}},\
  }\href {\doibase 10.1103/PhysRevD.32.1863} {\bibfield  {journal} {\bibinfo
  {journal} {Phys.Rev.}\ }\textbf {\bibinfo {volume} {D32}},\ \bibinfo {pages}
  {1863} (\bibinfo {year} {1985})}\BibitemShut {NoStop}%
%%CITATION = PHRVA,D32,1863;%%
\bibitem [{\citenamefont {Belgiorno}\ and\ \citenamefont
  {Martellini}(1999)}]{Belgiorno:1998gj}%
  \BibitemOpen
  \bibfield  {author} {\bibinfo {author} {\bibfnamefont {F.}~\bibnamefont
  {Belgiorno}}\ and\ \bibinfo {author} {\bibfnamefont {M.}~\bibnamefont
  {Martellini}},\ }\href {\doibase 10.1016/S0370-2693(99)00313-5} {\bibfield
  {journal} {\bibinfo  {journal} {Phys.Lett.}\ }\textbf {\bibinfo {volume}
  {B453}},\ \bibinfo {pages} {17} (\bibinfo {year} {1999})},\ \Eprint
  {http://arxiv.org/abs/gr-qc/9811060} {arXiv:gr-qc/9811060 [gr-qc]}
  \BibitemShut {NoStop}%
%%CITATION = GR-QC/9811060;%%
\bibitem [{\citenamefont {Finster}\ \emph {et~al.}(2000)\citenamefont
  {Finster}, \citenamefont {Kamran}, \citenamefont {Smoller},\ and\
  \citenamefont {Yau}}]{Finster:1999ry}%
  \BibitemOpen
  \bibfield  {author} {\bibinfo {author} {\bibfnamefont {F.}~\bibnamefont
  {Finster}}, \bibinfo {author} {\bibfnamefont {N.}~\bibnamefont {Kamran}},
  \bibinfo {author} {\bibfnamefont {J.}~\bibnamefont {Smoller}}, \ and\
  \bibinfo {author} {\bibfnamefont {S.-T.}\ \bibnamefont {Yau}},\ }\href@noop
  {} {\bibfield  {journal} {\bibinfo  {journal} {Commun.Pure Appl.Math.}\
  }\textbf {\bibinfo {volume} {53}},\ \bibinfo {pages} {902} (\bibinfo {year}
  {2000})},\ \Eprint {http://arxiv.org/abs/gr-qc/9905047} {arXiv:gr-qc/9905047
  [gr-qc]} \BibitemShut {NoStop}%
%%CITATION = GR-QC/9905047;%%
\bibitem [{\citenamefont {Finster}\ \emph {et~al.}(2003)\citenamefont
  {Finster}, \citenamefont {Kamran}, \citenamefont {Smoller},\ and\
  \citenamefont {Yau}}]{Finster:2000jz}%
  \BibitemOpen
  \bibfield  {author} {\bibinfo {author} {\bibfnamefont {F.}~\bibnamefont
  {Finster}}, \bibinfo {author} {\bibfnamefont {N.}~\bibnamefont {Kamran}},
  \bibinfo {author} {\bibfnamefont {J.}~\bibnamefont {Smoller}}, \ and\
  \bibinfo {author} {\bibfnamefont {S.-T.}\ \bibnamefont {Yau}},\ }\href@noop
  {} {\bibfield  {journal} {\bibinfo  {journal} {Adv.Theor.Math.Phys.}\
  }\textbf {\bibinfo {volume} {7}},\ \bibinfo {pages} {25} (\bibinfo {year}
  {2003})},\ \Eprint {http://arxiv.org/abs/gr-qc/0005088} {arXiv:gr-qc/0005088
  [gr-qc]} \BibitemShut {NoStop}%
%%CITATION = GR-QC/0005088;%%
\bibitem [{\citenamefont {Batic}\ \emph {et~al.}(2005)\citenamefont {Batic},
  \citenamefont {Schmid},\ and\ \citenamefont {Winklmeier}}]{Batic:2004sz}%
  \BibitemOpen
  \bibfield  {author} {\bibinfo {author} {\bibfnamefont {D.}~\bibnamefont
  {Batic}}, \bibinfo {author} {\bibfnamefont {H.}~\bibnamefont {Schmid}}, \
  and\ \bibinfo {author} {\bibfnamefont {M.}~\bibnamefont {Winklmeier}},\
  }\href {\doibase 10.1063/1.1818720} {\bibfield  {journal} {\bibinfo
  {journal} {J.Math.Phys.}\ }\textbf {\bibinfo {volume} {46}},\ \bibinfo
  {pages} {012504} (\bibinfo {year} {2005})},\ \Eprint
  {http://arxiv.org/abs/math-ph/0402047} {arXiv:math-ph/0402047 [math-ph]}
  \BibitemShut {NoStop}%
%%CITATION = MATH-PH/0402047;%%
\bibitem [{\citenamefont {Batic}\ and\ \citenamefont
  {Schmid}(2005)}]{Batic:2005va}%
  \BibitemOpen
  \bibfield  {author} {\bibinfo {author} {\bibfnamefont {D.}~\bibnamefont
  {Batic}}\ and\ \bibinfo {author} {\bibfnamefont {H.}~\bibnamefont {Schmid}},\
  }\href@noop {} {\  (\bibinfo {year} {2005})},\ \Eprint
  {http://arxiv.org/abs/gr-qc/0512112} {arXiv:gr-qc/0512112 [gr-qc]}
  \BibitemShut {NoStop}%
%%CITATION = GR-QC/0512112;%%
\bibitem [{\citenamefont {Batic}\ and\ \citenamefont
  {Schmid}(2006)}]{Batic:2006vs}%
  \BibitemOpen
  \bibfield  {author} {\bibinfo {author} {\bibfnamefont {D.}~\bibnamefont
  {Batic}}\ and\ \bibinfo {author} {\bibfnamefont {H.}~\bibnamefont {Schmid}},\
  }\href {\doibase 10.1143/PTP.116.517} {\bibfield  {journal} {\bibinfo
  {journal} {Prog.Theor.Phys.}\ }\textbf {\bibinfo {volume} {116}},\ \bibinfo
  {pages} {517} (\bibinfo {year} {2006})},\ \Eprint
  {http://arxiv.org/abs/gr-qc/0606050} {arXiv:gr-qc/0606050 [gr-qc]}
  \BibitemShut {NoStop}%
%%CITATION = GR-QC/0606050;%%
\bibitem [{\citenamefont {Winklmeier}\ and\ \citenamefont
  {Yamada}(2006)}]{Winklmeier:2006me}%
  \BibitemOpen
  \bibfield  {author} {\bibinfo {author} {\bibfnamefont {M.}~\bibnamefont
  {Winklmeier}}\ and\ \bibinfo {author} {\bibfnamefont {O.}~\bibnamefont
  {Yamada}},\ }\href {\doibase 10.1063/1.2358394} {\bibfield  {journal}
  {\bibinfo  {journal} {J.Math.Phys.}\ }\textbf {\bibinfo {volume} {47}},\
  \bibinfo {pages} {102503} (\bibinfo {year} {2006})},\ \Eprint
  {http://arxiv.org/abs/gr-qc/0605146} {arXiv:gr-qc/0605146 [gr-qc]}
  \BibitemShut {NoStop}%
%%CITATION = GR-QC/0605146;%%
\bibitem [{\citenamefont {He}\ and\ \citenamefont {Jing}(2006)}]{He:2006jv}%
  \BibitemOpen
  \bibfield  {author} {\bibinfo {author} {\bibfnamefont {X.}~\bibnamefont
  {He}}\ and\ \bibinfo {author} {\bibfnamefont {J.}~\bibnamefont {Jing}},\
  }\href {\doibase 10.1016/j.nuclphysb.2006.08.015} {\bibfield  {journal}
  {\bibinfo  {journal} {Nucl.Phys.}\ }\textbf {\bibinfo {volume} {B755}},\
  \bibinfo {pages} {313} (\bibinfo {year} {2006})},\ \Eprint
  {http://arxiv.org/abs/gr-qc/0611003} {arXiv:gr-qc/0611003 [gr-qc]}
  \BibitemShut {NoStop}%
%%CITATION = GR-QC/0611003;%%
\bibitem [{\citenamefont {Zhou}\ and\ \citenamefont
  {Liu}(2008)}]{Zhou:2008zzf}%
  \BibitemOpen
  \bibfield  {author} {\bibinfo {author} {\bibfnamefont {S.}~\bibnamefont
  {Zhou}}\ and\ \bibinfo {author} {\bibfnamefont {W.}~\bibnamefont {Liu}},\
  }\href {\doibase 10.1103/PhysRevD.77.104021} {\bibfield  {journal} {\bibinfo
  {journal} {Phys.Rev.}\ }\textbf {\bibinfo {volume} {D77}},\ \bibinfo {pages}
  {104021} (\bibinfo {year} {2008})}\BibitemShut {NoStop}%
%%CITATION = PHRVA,D77,104021;%%
\bibitem [{\citenamefont {Dolan}\ and\ \citenamefont
  {Gair}(2009)}]{Dolan:2009kj}%
  \BibitemOpen
  \bibfield  {author} {\bibinfo {author} {\bibfnamefont {S.}~\bibnamefont
  {Dolan}}\ and\ \bibinfo {author} {\bibfnamefont {J.}~\bibnamefont {Gair}},\
  }\href {\doibase 10.1088/0264-9381/26/17/175020} {\bibfield  {journal}
  {\bibinfo  {journal} {Class.Quant.Grav.}\ }\textbf {\bibinfo {volume} {26}},\
  \bibinfo {pages} {175020} (\bibinfo {year} {2009})},\ \Eprint
  {http://arxiv.org/abs/0905.2974} {arXiv:0905.2974 [gr-qc]} \BibitemShut
  {NoStop}%
%%CITATION = ARXIV:0905.2974;%%
\bibitem [{\citenamefont {Belgiorno}\ and\ \citenamefont
  {Cacciatori}(2010)}]{Belgiorno:2008hk}%
  \BibitemOpen
  \bibfield  {author} {\bibinfo {author} {\bibfnamefont {F.}~\bibnamefont
  {Belgiorno}}\ and\ \bibinfo {author} {\bibfnamefont {S.~L.}\ \bibnamefont
  {Cacciatori}},\ }\href {\doibase 10.1063/1.3300401} {\bibfield  {journal}
  {\bibinfo  {journal} {J.Math.Phys.}\ }\textbf {\bibinfo {volume} {51}},\
  \bibinfo {pages} {033517} (\bibinfo {year} {2010})},\ \Eprint
  {http://arxiv.org/abs/0803.2496} {arXiv:0803.2496 [math-ph]} \BibitemShut
  {NoStop}%
%%CITATION = ARXIV:0803.2496;%%
\bibitem [{\citenamefont {Leaver}(1985)}]{Leaver:1985ax}%
  \BibitemOpen
  \bibfield  {author} {\bibinfo {author} {\bibfnamefont {E.}~\bibnamefont
  {Leaver}},\ }\href@noop {} {\bibfield  {journal} {\bibinfo  {journal}
  {Proc.Roy.Soc.Lond.}\ }\textbf {\bibinfo {volume} {A402}},\ \bibinfo {pages}
  {285} (\bibinfo {year} {1985})}\BibitemShut {NoStop}%
%%CITATION = PRSLA,A402,285;%%
\bibitem [{\citenamefont {Cho}\ \emph {et~al.}(2012)\citenamefont {Cho},
  \citenamefont {Cornell}, \citenamefont {Doukas}, \citenamefont {Huang},\ and\
  \citenamefont {Naylor}}]{Cho:2011sf}%
  \BibitemOpen
  \bibfield  {author} {\bibinfo {author} {\bibfnamefont {H.}~\bibnamefont
  {Cho}}, \bibinfo {author} {\bibfnamefont {A.}~\bibnamefont {Cornell}},
  \bibinfo {author} {\bibfnamefont {J.}~\bibnamefont {Doukas}}, \bibinfo
  {author} {\bibfnamefont {T.}~\bibnamefont {Huang}}, \ and\ \bibinfo {author}
  {\bibfnamefont {W.}~\bibnamefont {Naylor}},\ }\href {\doibase
  10.1155/2012/281705} {\bibfield  {journal} {\bibinfo  {journal}
  {Adv.Math.Phys.}\ }\textbf {\bibinfo {volume} {2012}},\ \bibinfo {pages}
  {281705} (\bibinfo {year} {2012})},\ \Eprint {http://arxiv.org/abs/1111.5024}
  {arXiv:1111.5024 [gr-qc]} \BibitemShut {NoStop}%
%%CITATION = ARXIV:1111.5024;%%
\bibitem [{\citenamefont {Dudley}\ and\ \citenamefont
  {Finley}(1977)}]{Dudley:1977zz}%
  \BibitemOpen
  \bibfield  {author} {\bibinfo {author} {\bibfnamefont {A.~L.}\ \bibnamefont
  {Dudley}}\ and\ \bibinfo {author} {\bibfnamefont {J.}~\bibnamefont
  {Finley}},\ }\href {\doibase 10.1103/PhysRevLett.38.1505} {\bibfield
  {journal} {\bibinfo  {journal} {Phys.Rev.Lett.}\ }\textbf {\bibinfo {volume}
  {38}},\ \bibinfo {pages} {1505} (\bibinfo {year} {1977})}\BibitemShut
  {NoStop}%
%%CITATION = PRLTA,38,1505;%%
\bibitem [{\citenamefont {Dudley}\ and\ \citenamefont
  {Finley}(1979)}]{Dudley:1978vd}%
  \BibitemOpen
  \bibfield  {author} {\bibinfo {author} {\bibfnamefont {A.~L.}\ \bibnamefont
  {Dudley}}\ and\ \bibinfo {author} {\bibfnamefont {I.}~\bibnamefont {Finley},
  \bibfnamefont {J.D.}},\ }\href {\doibase 10.1063/1.524064} {\bibfield
  {journal} {\bibinfo  {journal} {J.Math.Phys.}\ }\textbf {\bibinfo {volume}
  {20}},\ \bibinfo {pages} {311} (\bibinfo {year} {1979})}\BibitemShut
  {NoStop}%
%%CITATION = JMAPA,20,311;%%
\bibitem [{\citenamefont {Bellezza}\ and\ \citenamefont
  {Ferrari}(1984)}]{Bellezza:1984}%
  \BibitemOpen
  \bibfield  {author} {\bibinfo {author} {\bibfnamefont {V.}~\bibnamefont
  {Bellezza}}\ and\ \bibinfo {author} {\bibfnamefont {V.}~\bibnamefont
  {Ferrari}},\ }\href@noop {} {\bibfield  {journal} {\bibinfo  {journal}
  {J.Math.Phys.}\ }\textbf {\bibinfo {volume} {25}},\ \bibinfo {pages} {1985}
  (\bibinfo {year} {1984})}\BibitemShut {NoStop}%
\bibitem [{\citenamefont {Chandrasekhar}(1983)}]{Chandra}%
  \BibitemOpen
  \bibfield  {author} {\bibinfo {author} {\bibfnamefont {S.}~\bibnamefont
  {Chandrasekhar}},\ }\href@noop {} {\emph {\bibinfo {title} {{The mathematical
  theory of black holes}}}}\ (\bibinfo {year} {1983})\BibitemShut {NoStop}%
%%CITATION = INSPIRE-224457;%%
\bibitem [{\citenamefont {Mashhoon}(1985)}]{Mashhoon:1985}%
  \BibitemOpen
  \bibfield  {author} {\bibinfo {author} {\bibfnamefont {B.}~\bibnamefont
  {Mashhoon}},\ }\href {\doibase 10.1103/PhysRevD.31.290} {\bibfield  {journal}
  {\bibinfo  {journal} {Phys. Rev. D}\ }\textbf {\bibinfo {volume} {31}},\
  \bibinfo {pages} {290} (\bibinfo {year} {1985})}\BibitemShut {NoStop}%
\bibitem [{\citenamefont {Goebel}(1972)}]{1972ApJ...172L..95G}%
  \BibitemOpen
  \bibfield  {author} {\bibinfo {author} {\bibfnamefont {C.}~\bibnamefont
  {Goebel}},\ }\href {\doibase 10.1086/180898} {\bibfield  {journal} {\bibinfo
  {journal} {Astrophys.J.Lett.}\ }\textbf {\bibinfo {volume} {172}},\ \bibinfo
  {pages} {L95} (\bibinfo {year} {1972})}\BibitemShut {NoStop}%
\bibitem [{\citenamefont {Cardoso}\ \emph {et~al.}(2009)\citenamefont
  {Cardoso}, \citenamefont {Miranda}, \citenamefont {Berti}, \citenamefont
  {Witek},\ and\ \citenamefont {Zanchin}}]{Cardoso:2008bp}%
  \BibitemOpen
  \bibfield  {author} {\bibinfo {author} {\bibfnamefont {V.}~\bibnamefont
  {Cardoso}}, \bibinfo {author} {\bibfnamefont {A.~S.}\ \bibnamefont
  {Miranda}}, \bibinfo {author} {\bibfnamefont {E.}~\bibnamefont {Berti}},
  \bibinfo {author} {\bibfnamefont {H.}~\bibnamefont {Witek}}, \ and\ \bibinfo
  {author} {\bibfnamefont {V.~T.}\ \bibnamefont {Zanchin}},\ }\href {\doibase
  10.1103/PhysRevD.79.064016} {\bibfield  {journal} {\bibinfo  {journal}
  {Phys.Rev.}\ }\textbf {\bibinfo {volume} {D79}},\ \bibinfo {pages} {064016}
  (\bibinfo {year} {2009})},\ \Eprint {http://arxiv.org/abs/0812.1806}
  {arXiv:0812.1806 [hep-th]} \BibitemShut {NoStop}%
%%CITATION = ARXIV:0812.1806;%%
\bibitem [{\citenamefont {Plebanski}\ and\ \citenamefont
  {Demianski}(1976)}]{Plebanski:1976gy}%
  \BibitemOpen
  \bibfield  {author} {\bibinfo {author} {\bibfnamefont {J.}~\bibnamefont
  {Plebanski}}\ and\ \bibinfo {author} {\bibfnamefont {M.}~\bibnamefont
  {Demianski}},\ }\href {\doibase 10.1016/0003-4916(76)90240-2} {\bibfield
  {journal} {\bibinfo  {journal} {Annals Phys.}\ }\textbf {\bibinfo {volume}
  {98}},\ \bibinfo {pages} {98} (\bibinfo {year} {1976})}\BibitemShut {NoStop}%
%%CITATION = APNYA,98,98;%%
\bibitem [{\citenamefont {Kokkotas}(1993)}]{Kokkotas:1993}%
  \BibitemOpen
  \bibfield  {author} {\bibinfo {author} {\bibfnamefont {K.~D.}\ \bibnamefont
  {Kokkotas}},\ }\href@noop {} {\bibfield  {journal} {\bibinfo  {journal}
  {Nuovo Cimento}\ }\textbf {\bibinfo {volume} {108}},\ \bibinfo {pages} {991}
  (\bibinfo {year} {1993})}\BibitemShut {NoStop}%
\bibitem [{\citenamefont {Pani}\ \emph
  {et~al.}(2012{\natexlab{a}})\citenamefont {Pani}, \citenamefont {Cardoso},
  \citenamefont {Gualtieri}, \citenamefont {Berti},\ and\ \citenamefont
  {Ishibashi}}]{Pani:2012vp}%
  \BibitemOpen
  \bibfield  {author} {\bibinfo {author} {\bibfnamefont {P.}~\bibnamefont
  {Pani}}, \bibinfo {author} {\bibfnamefont {V.}~\bibnamefont {Cardoso}},
  \bibinfo {author} {\bibfnamefont {L.}~\bibnamefont {Gualtieri}}, \bibinfo
  {author} {\bibfnamefont {E.}~\bibnamefont {Berti}}, \ and\ \bibinfo {author}
  {\bibfnamefont {A.}~\bibnamefont {Ishibashi}},\ }\href {\doibase
  10.1103/PhysRevLett.109.131102} {\bibfield  {journal} {\bibinfo  {journal}
  {Phys.Rev.Lett.}\ }\textbf {\bibinfo {volume} {109}},\ \bibinfo {pages}
  {131102} (\bibinfo {year} {2012}{\natexlab{a}})},\ \Eprint
  {http://arxiv.org/abs/1209.0465} {arXiv:1209.0465 [gr-qc]} \BibitemShut
  {NoStop}%
%%CITATION = ARXIV:1209.0465;%%
\bibitem [{\citenamefont {Pani}\ \emph
  {et~al.}(2012{\natexlab{b}})\citenamefont {Pani}, \citenamefont {Cardoso},
  \citenamefont {Gualtieri}, \citenamefont {Berti},\ and\ \citenamefont
  {Ishibashi}}]{Pani:2012bp}%
  \BibitemOpen
  \bibfield  {author} {\bibinfo {author} {\bibfnamefont {P.}~\bibnamefont
  {Pani}}, \bibinfo {author} {\bibfnamefont {V.}~\bibnamefont {Cardoso}},
  \bibinfo {author} {\bibfnamefont {L.}~\bibnamefont {Gualtieri}}, \bibinfo
  {author} {\bibfnamefont {E.}~\bibnamefont {Berti}}, \ and\ \bibinfo {author}
  {\bibfnamefont {A.}~\bibnamefont {Ishibashi}},\ }\href {\doibase
  10.1103/PhysRevD.86.104017} {\bibfield  {journal} {\bibinfo  {journal}
  {Phys.Rev.}\ }\textbf {\bibinfo {volume} {D86}},\ \bibinfo {pages} {104017}
  (\bibinfo {year} {2012}{\natexlab{b}})},\ \Eprint
  {http://arxiv.org/abs/1209.0773} {arXiv:1209.0773 [gr-qc]} \BibitemShut
  {NoStop}%
%%CITATION = ARXIV:1209.0773;%%
\bibitem [{\citenamefont {Chandrasekhar}\ and\ \citenamefont
  {Ferrari}(1991)}]{ChandraFerrari91}%
  \BibitemOpen
  \bibfield  {author} {\bibinfo {author} {\bibfnamefont {S.}~\bibnamefont
  {Chandrasekhar}}\ and\ \bibinfo {author} {\bibfnamefont {V.}~\bibnamefont
  {Ferrari}},\ }\href@noop {} {\bibfield  {journal} {\bibinfo  {journal}
  {Proc.Roy.Soc.Lond.}\ }\textbf {\bibinfo {volume} {A433}},\ \bibinfo {pages}
  {423} (\bibinfo {year} {1991})}\BibitemShut {NoStop}%
\bibitem [{\citenamefont {Kojima}(1992)}]{Kojima:1992ie}%
  \BibitemOpen
  \bibfield  {author} {\bibinfo {author} {\bibfnamefont {Y.}~\bibnamefont
  {Kojima}},\ }\href {\doibase 10.1103/PhysRevD.46.4289} {\bibfield  {journal}
  {\bibinfo  {journal} {Phys.Rev.}\ }\textbf {\bibinfo {volume} {D46}},\
  \bibinfo {pages} {4289} (\bibinfo {year} {1992})}\BibitemShut {NoStop}%
%%CITATION = PHRVA,D46,4289;%%
\bibitem [{\citenamefont {{Kojima}}(1993{\natexlab{a}})}]{1993ApJ...414..247K}%
  \BibitemOpen
  \bibfield  {author} {\bibinfo {author} {\bibfnamefont {Y.}~\bibnamefont
  {{Kojima}}},\ }\href {\doibase 10.1086/173073} {\bibfield  {journal}
  {\bibinfo  {journal} {Astrophys.J.}\ }\textbf {\bibinfo {volume} {414}},\
  \bibinfo {pages} {247} (\bibinfo {year} {1993}{\natexlab{a}})}\BibitemShut
  {NoStop}%
\bibitem [{\citenamefont {Ferrari}\ \emph {et~al.}(2007)\citenamefont
  {Ferrari}, \citenamefont {Gualtieri},\ and\ \citenamefont
  {Marassi}}]{Ferrari:2007rc}%
  \BibitemOpen
  \bibfield  {author} {\bibinfo {author} {\bibfnamefont {V.}~\bibnamefont
  {Ferrari}}, \bibinfo {author} {\bibfnamefont {L.}~\bibnamefont {Gualtieri}},
  \ and\ \bibinfo {author} {\bibfnamefont {S.}~\bibnamefont {Marassi}},\ }\href
  {\doibase 10.1103/PhysRevD.76.104033} {\bibfield  {journal} {\bibinfo
  {journal} {Phys.Rev.}\ }\textbf {\bibinfo {volume} {D76}},\ \bibinfo {pages}
  {104033} (\bibinfo {year} {2007})},\ \Eprint {http://arxiv.org/abs/0709.2925}
  {arXiv:0709.2925 [gr-qc]} \BibitemShut {NoStop}%
%%CITATION = ARXIV:0709.2925;%%
\bibitem [{web()}]{webpage}%
  \BibitemOpen
  \href@noop {} {\ }\bibinfo {note}
  {\noindent\url{http://blackholes.ist.utl.pt/?page=Files},\\
  \url{http://www.phy.olemiss.edu/~berti/qnms.html}}\BibitemShut {NoStop}%
\bibitem [{\citenamefont {Pani}(2013)}]{Pani:2013pma}%
  \BibitemOpen
  \bibfield  {author} {\bibinfo {author} {\bibfnamefont {P.}~\bibnamefont
  {Pani}},\ }\href@noop {} {\  (\bibinfo {year} {2013})},\ \Eprint
  {http://arxiv.org/abs/1305.6759} {arXiv:1305.6759 [gr-qc]} \BibitemShut
  {NoStop}%
%%CITATION = ARXIV:1305.6759;%%
\bibitem [{\citenamefont {Berti}\ \emph {et~al.}(2006)\citenamefont {Berti},
  \citenamefont {Cardoso},\ and\ \citenamefont {Will}}]{Berti:2005ys}%
  \BibitemOpen
  \bibfield  {author} {\bibinfo {author} {\bibfnamefont {E.}~\bibnamefont
  {Berti}}, \bibinfo {author} {\bibfnamefont {V.}~\bibnamefont {Cardoso}}, \
  and\ \bibinfo {author} {\bibfnamefont {C.~M.}\ \bibnamefont {Will}},\ }\href
  {\doibase 10.1103/PhysRevD.73.064030} {\bibfield  {journal} {\bibinfo
  {journal} {Phys.Rev.}\ }\textbf {\bibinfo {volume} {D73}},\ \bibinfo {pages}
  {064030} (\bibinfo {year} {2006})},\ \Eprint
  {http://arxiv.org/abs/gr-qc/0512160} {arXiv:gr-qc/0512160 [gr-qc]}
  \BibitemShut {NoStop}%
%%CITATION = GR-QC/0512160;%%
\bibitem [{\citenamefont {Zerilli}(1974)}]{Zerilli:1974ai}%
  \BibitemOpen
  \bibfield  {author} {\bibinfo {author} {\bibfnamefont {F.}~\bibnamefont
  {Zerilli}},\ }\href {\doibase 10.1103/PhysRevD.9.860} {\bibfield  {journal}
  {\bibinfo  {journal} {Phys.Rev.}\ }\textbf {\bibinfo {volume} {D9}},\
  \bibinfo {pages} {860} (\bibinfo {year} {1974})}\BibitemShut {NoStop}%
%%CITATION = PHRVA,D9,860;%%
\bibitem [{\citenamefont {Leaver}(1990)}]{Leaver:1990zz}%
  \BibitemOpen
  \bibfield  {author} {\bibinfo {author} {\bibfnamefont {E.~W.}\ \bibnamefont
  {Leaver}},\ }\href {\doibase 10.1103/PhysRevD.41.2986} {\bibfield  {journal}
  {\bibinfo  {journal} {Phys.Rev.}\ }\textbf {\bibinfo {volume} {D41}},\
  \bibinfo {pages} {2986} (\bibinfo {year} {1990})}\BibitemShut {NoStop}%
%%CITATION = PHRVA,D41,2986;%%
\bibitem [{\citenamefont {Rosa}\ and\ \citenamefont
  {Dolan}(2012)}]{Rosa:2011my}%
  \BibitemOpen
  \bibfield  {author} {\bibinfo {author} {\bibfnamefont {J.~G.}\ \bibnamefont
  {Rosa}}\ and\ \bibinfo {author} {\bibfnamefont {S.~R.}\ \bibnamefont
  {Dolan}},\ }\href {\doibase 10.1103/PhysRevD.85.044043} {\bibfield  {journal}
  {\bibinfo  {journal} {Phys.Rev.}\ }\textbf {\bibinfo {volume} {D85}},\
  \bibinfo {pages} {044043} (\bibinfo {year} {2012})},\ \Eprint
  {http://arxiv.org/abs/1110.4494} {arXiv:1110.4494 [hep-th]} \BibitemShut
  {NoStop}%
%%CITATION = ARXIV:1110.4494;%%
\bibitem [{\citenamefont {{Kojima}}(1993{\natexlab{b}})}]{1993PThPh..90..977K}%
  \BibitemOpen
  \bibfield  {author} {\bibinfo {author} {\bibfnamefont {Y.}~\bibnamefont
  {{Kojima}}},\ }\href {\doibase 10.1143/PTP.90.977} {\bibfield  {journal}
  {\bibinfo  {journal} {Progress of Theoretical Physics}\ }\textbf {\bibinfo
  {volume} {90}},\ \bibinfo {pages} {977} (\bibinfo {year}
  {1993}{\natexlab{b}})}\BibitemShut {NoStop}%
\bibitem [{\citenamefont {Berti}(2004)}]{Berti:2004md}%
  \BibitemOpen
  \bibfield  {author} {\bibinfo {author} {\bibfnamefont {E.}~\bibnamefont
  {Berti}},\ }\href@noop {} {\bibfield  {journal} {\bibinfo  {journal}
  {Conf.Proc.}\ }\textbf {\bibinfo {volume} {C0405132}},\ \bibinfo {pages}
  {145} (\bibinfo {year} {2004})},\ \Eprint
  {http://arxiv.org/abs/gr-qc/0411025} {arXiv:gr-qc/0411025 [gr-qc]}
  \BibitemShut {NoStop}%
%%CITATION = GR-QC/0411025;%%
\bibitem [{\citenamefont {Mellor}\ and\ \citenamefont
  {Moss}(1990)}]{Mellor:1989ac}%
  \BibitemOpen
  \bibfield  {author} {\bibinfo {author} {\bibfnamefont {F.}~\bibnamefont
  {Mellor}}\ and\ \bibinfo {author} {\bibfnamefont {I.}~\bibnamefont {Moss}},\
  }\href {\doibase 10.1103/PhysRevD.41.403} {\bibfield  {journal} {\bibinfo
  {journal} {Phys.Rev.}\ }\textbf {\bibinfo {volume} {D41}},\ \bibinfo {pages}
  {403} (\bibinfo {year} {1990})}\BibitemShut {NoStop}%
%%CITATION = PHRVA,D41,403;%%
\bibitem [{\citenamefont {Cardoso}\ and\ \citenamefont
  {Lemos}(2001)}]{Cardoso:2001bb}%
  \BibitemOpen
  \bibfield  {author} {\bibinfo {author} {\bibfnamefont {V.}~\bibnamefont
  {Cardoso}}\ and\ \bibinfo {author} {\bibfnamefont {J.~P.}\ \bibnamefont
  {Lemos}},\ }\href {\doibase 10.1103/PhysRevD.64.084017} {\bibfield  {journal}
  {\bibinfo  {journal} {Phys.Rev.}\ }\textbf {\bibinfo {volume} {D64}},\
  \bibinfo {pages} {084017} (\bibinfo {year} {2001})},\ \Eprint
  {http://arxiv.org/abs/gr-qc/0105103} {arXiv:gr-qc/0105103 [gr-qc]}
  \BibitemShut {NoStop}%
%%CITATION = GR-QC/0105103;%%
\bibitem [{\citenamefont {Berti}\ and\ \citenamefont
  {Kokkotas}(2003)}]{Berti:2003ud}%
  \BibitemOpen
  \bibfield  {author} {\bibinfo {author} {\bibfnamefont {E.}~\bibnamefont
  {Berti}}\ and\ \bibinfo {author} {\bibfnamefont {K.}~\bibnamefont
  {Kokkotas}},\ }\href {\doibase 10.1103/PhysRevD.67.064020} {\bibfield
  {journal} {\bibinfo  {journal} {Phys.Rev.}\ }\textbf {\bibinfo {volume}
  {D67}},\ \bibinfo {pages} {064020} (\bibinfo {year} {2003})},\ \Eprint
  {http://arxiv.org/abs/gr-qc/0301052} {arXiv:gr-qc/0301052 [gr-qc]}
  \BibitemShut {NoStop}%
%%CITATION = GR-QC/0301052;%%
\bibitem [{\citenamefont {Cardoso}\ and\ \citenamefont
  {Gualtieri}(2009)}]{Cardoso:2009pk}%
  \BibitemOpen
  \bibfield  {author} {\bibinfo {author} {\bibfnamefont {V.}~\bibnamefont
  {Cardoso}}\ and\ \bibinfo {author} {\bibfnamefont {L.}~\bibnamefont
  {Gualtieri}},\ }\href {\doibase 10.1103/PhysRevD.80.064008,
  10.1103/PhysRevD.81.089903} {\bibfield  {journal} {\bibinfo  {journal}
  {Phys.Rev.}\ }\textbf {\bibinfo {volume} {D80}},\ \bibinfo {pages} {064008}
  (\bibinfo {year} {2009})},\ \Eprint {http://arxiv.org/abs/0907.5008}
  {arXiv:0907.5008 [gr-qc]} \BibitemShut {NoStop}%
%%CITATION = ARXIV:0907.5008;%%
\bibitem [{\citenamefont {Molina}\ \emph {et~al.}(2010)\citenamefont {Molina},
  \citenamefont {Pani}, \citenamefont {Cardoso},\ and\ \citenamefont
  {Gualtieri}}]{Molina:2010fb}%
  \BibitemOpen
  \bibfield  {author} {\bibinfo {author} {\bibfnamefont {C.}~\bibnamefont
  {Molina}}, \bibinfo {author} {\bibfnamefont {P.}~\bibnamefont {Pani}},
  \bibinfo {author} {\bibfnamefont {V.}~\bibnamefont {Cardoso}}, \ and\
  \bibinfo {author} {\bibfnamefont {L.}~\bibnamefont {Gualtieri}},\ }\href
  {\doibase 10.1103/PhysRevD.81.124021} {\bibfield  {journal} {\bibinfo
  {journal} {Phys.Rev.}\ }\textbf {\bibinfo {volume} {D81}},\ \bibinfo {pages}
  {124021} (\bibinfo {year} {2010})},\ \Eprint {http://arxiv.org/abs/1004.4007}
  {arXiv:1004.4007 [gr-qc]} \BibitemShut {NoStop}%
%%CITATION = ARXIV:1004.4007;%%
\bibitem [{\citenamefont {Dorband}\ \emph {et~al.}(2006)\citenamefont
  {Dorband}, \citenamefont {Berti}, \citenamefont {Diener}, \citenamefont
  {Schnetter},\ and\ \citenamefont {Tiglio}}]{Dorband:2006gg}%
  \BibitemOpen
  \bibfield  {author} {\bibinfo {author} {\bibfnamefont {E.~N.}\ \bibnamefont
  {Dorband}}, \bibinfo {author} {\bibfnamefont {E.}~\bibnamefont {Berti}},
  \bibinfo {author} {\bibfnamefont {P.}~\bibnamefont {Diener}}, \bibinfo
  {author} {\bibfnamefont {E.}~\bibnamefont {Schnetter}}, \ and\ \bibinfo
  {author} {\bibfnamefont {M.}~\bibnamefont {Tiglio}},\ }\href {\doibase
  10.1103/PhysRevD.74.084028} {\bibfield  {journal} {\bibinfo  {journal}
  {Phys.Rev.}\ }\textbf {\bibinfo {volume} {D74}},\ \bibinfo {pages} {084028}
  (\bibinfo {year} {2006})},\ \Eprint {http://arxiv.org/abs/gr-qc/0608091}
  {arXiv:gr-qc/0608091 [gr-qc]} \BibitemShut {NoStop}%
%%CITATION = GR-QC/0608091;%%
\bibitem [{\citenamefont {Witek}\ \emph {et~al.}(2012)\citenamefont {Witek},
  \citenamefont {Cardoso}, \citenamefont {Ishibashi},\ and\ \citenamefont
  {Sperhake}}]{Witek:2012tr}%
  \BibitemOpen
  \bibfield  {author} {\bibinfo {author} {\bibfnamefont {H.}~\bibnamefont
  {Witek}}, \bibinfo {author} {\bibfnamefont {V.}~\bibnamefont {Cardoso}},
  \bibinfo {author} {\bibfnamefont {A.}~\bibnamefont {Ishibashi}}, \ and\
  \bibinfo {author} {\bibfnamefont {U.}~\bibnamefont {Sperhake}},\ }\href@noop
  {} {\  (\bibinfo {year} {2012})},\ \Eprint {http://arxiv.org/abs/1212.0551}
  {arXiv:1212.0551 [gr-qc]} \BibitemShut {NoStop}%
%%CITATION = ARXIV:1212.0551;%%
\bibitem [{\citenamefont {Dolan}(2012)}]{Dolan:2012yt}%
  \BibitemOpen
  \bibfield  {author} {\bibinfo {author} {\bibfnamefont {S.~R.}\ \bibnamefont
  {Dolan}},\ }\href@noop {} {\  (\bibinfo {year} {2012})},\ \Eprint
  {http://arxiv.org/abs/1212.1477} {arXiv:1212.1477 [gr-qc]} \BibitemShut
  {NoStop}%
%%CITATION = ARXIV:1212.1477;%%
\bibitem [{\citenamefont {{Leung}}\ \emph {et~al.}(1999)\citenamefont
  {{Leung}}, \citenamefont {{Maassen van den Brink}}, \citenamefont {{Suen}},
  \citenamefont {{Wong}},\ and\ \citenamefont {{Young}}}]{1999math.ph...9030L}%
  \BibitemOpen
  \bibfield  {author} {\bibinfo {author} {\bibfnamefont {P.~T.}\ \bibnamefont
  {{Leung}}}, \bibinfo {author} {\bibfnamefont {A.}~\bibnamefont {{Maassen van
  den Brink}}}, \bibinfo {author} {\bibfnamefont {W.~M.}\ \bibnamefont
  {{Suen}}}, \bibinfo {author} {\bibfnamefont {C.~W.}\ \bibnamefont {{Wong}}},
  \ and\ \bibinfo {author} {\bibfnamefont {K.}~\bibnamefont {{Young}}},\
  }\href@noop {} {\bibfield  {journal} {\bibinfo  {journal} {ArXiv Mathematical
  Physics e-prints}\ } (\bibinfo {year} {1999})},\ \Eprint
  {http://arxiv.org/abs/arXiv:math-ph/9909030} {arXiv:math-ph/9909030}
  \BibitemShut {NoStop}%
\bibitem [{\citenamefont {Chen}\ and\ \citenamefont {Chu}(2010)}]{Chen:2010i}%
  \BibitemOpen
  \bibfield  {author} {\bibinfo {author} {\bibfnamefont {B.}~\bibnamefont
  {Chen}}\ and\ \bibinfo {author} {\bibfnamefont {C.-S.}\ \bibnamefont {Chu}},\
  }\href {\doibase 10.1007/JHEP05(2010)004} {\bibfield  {journal} {\bibinfo
  {journal} {JHEP}\ }\textbf {\bibinfo {volume} {1005}},\ \bibinfo {pages}
  {004} (\bibinfo {year} {2010})},\ \Eprint {http://arxiv.org/abs/1001.3208}
  {arXiv:1001.3208 [hep-th]} \BibitemShut {NoStop}%
%%CITATION = ARXIV:1001.3208;%%
\bibitem [{\citenamefont {Wang}\ \emph {et~al.}(2000)\citenamefont {Wang},
  \citenamefont {Lin},\ and\ \citenamefont {Abdalla}}]{Wang:2000gsa}%
  \BibitemOpen
  \bibfield  {author} {\bibinfo {author} {\bibfnamefont {B.}~\bibnamefont
  {Wang}}, \bibinfo {author} {\bibfnamefont {C.-Y.}\ \bibnamefont {Lin}}, \
  and\ \bibinfo {author} {\bibfnamefont {E.}~\bibnamefont {Abdalla}},\ }\href
  {\doibase 10.1016/S0370-2693(00)00409-3} {\bibfield  {journal} {\bibinfo
  {journal} {Phys.Lett.}\ }\textbf {\bibinfo {volume} {B481}},\ \bibinfo
  {pages} {79} (\bibinfo {year} {2000})},\ \Eprint
  {http://arxiv.org/abs/hep-th/0003295} {arXiv:hep-th/0003295 [hep-th]}
  \BibitemShut {NoStop}%
%%CITATION = HEP-TH/0003295;%%
\bibitem [{\citenamefont {Maldacena}(1998)}]{Maldacena:1997re}%
  \BibitemOpen
  \bibfield  {author} {\bibinfo {author} {\bibfnamefont {J.~M.}\ \bibnamefont
  {Maldacena}},\ }\href@noop {} {\bibfield  {journal} {\bibinfo  {journal}
  {Adv.Theor.Math.Phys.}\ }\textbf {\bibinfo {volume} {2}},\ \bibinfo {pages}
  {231} (\bibinfo {year} {1998})},\ \Eprint
  {http://arxiv.org/abs/hep-th/9711200} {arXiv:hep-th/9711200 [hep-th]}
  \BibitemShut {NoStop}%
%%CITATION = HEP-TH/9711200;%%
\bibitem [{\citenamefont {Caldarelli}\ \emph {et~al.}(2000)\citenamefont
  {Caldarelli}, \citenamefont {Cognola},\ and\ \citenamefont
  {Klemm}}]{Caldarelli:1999xj}%
  \BibitemOpen
  \bibfield  {author} {\bibinfo {author} {\bibfnamefont {M.~M.}\ \bibnamefont
  {Caldarelli}}, \bibinfo {author} {\bibfnamefont {G.}~\bibnamefont {Cognola}},
  \ and\ \bibinfo {author} {\bibfnamefont {D.}~\bibnamefont {Klemm}},\ }\href
  {\doibase 10.1088/0264-9381/17/2/310} {\bibfield  {journal} {\bibinfo
  {journal} {Class.Quant.Grav.}\ }\textbf {\bibinfo {volume} {17}},\ \bibinfo
  {pages} {399} (\bibinfo {year} {2000})},\ \Eprint
  {http://arxiv.org/abs/hep-th/9908022} {arXiv:hep-th/9908022 [hep-th]}
  \BibitemShut {NoStop}%
%%CITATION = HEP-TH/9908022;%%
\bibitem [{\citenamefont {Hawking}\ and\ \citenamefont
  {Reall}(2000)}]{Hawking:1999dp}%
  \BibitemOpen
  \bibfield  {author} {\bibinfo {author} {\bibfnamefont {S.}~\bibnamefont
  {Hawking}}\ and\ \bibinfo {author} {\bibfnamefont {H.}~\bibnamefont
  {Reall}},\ }\href {\doibase 10.1103/PhysRevD.61.024014} {\bibfield  {journal}
  {\bibinfo  {journal} {Phys.Rev.}\ }\textbf {\bibinfo {volume} {D61}},\
  \bibinfo {pages} {024014} (\bibinfo {year} {2000})},\ \Eprint
  {http://arxiv.org/abs/hep-th/9908109} {arXiv:hep-th/9908109 [hep-th]}
  \BibitemShut {NoStop}%
%%CITATION = HEP-TH/9908109;%%
\bibitem [{\citenamefont {Kostelecky}\ and\ \citenamefont
  {Perry}(1996)}]{Kostelecky:1995ei}%
  \BibitemOpen
  \bibfield  {author} {\bibinfo {author} {\bibfnamefont {V.~A.}\ \bibnamefont
  {Kostelecky}}\ and\ \bibinfo {author} {\bibfnamefont {M.~J.}\ \bibnamefont
  {Perry}},\ }\href {\doibase 10.1016/0370-2693(95)01607-4} {\bibfield
  {journal} {\bibinfo  {journal} {Phys.Lett.}\ }\textbf {\bibinfo {volume}
  {B371}},\ \bibinfo {pages} {191} (\bibinfo {year} {1996})},\ \Eprint
  {http://arxiv.org/abs/hep-th/9512222} {arXiv:hep-th/9512222 [hep-th]}
  \BibitemShut {NoStop}%
%%CITATION = HEP-TH/9512222;%%
\bibitem [{\citenamefont {Caldarelli}\ and\ \citenamefont
  {Klemm}(1999)}]{Caldarelli:1998hg}%
  \BibitemOpen
  \bibfield  {author} {\bibinfo {author} {\bibfnamefont {M.~M.}\ \bibnamefont
  {Caldarelli}}\ and\ \bibinfo {author} {\bibfnamefont {D.}~\bibnamefont
  {Klemm}},\ }\href {\doibase 10.1016/S0550-3213(98)00846-3} {\bibfield
  {journal} {\bibinfo  {journal} {Nucl.Phys.}\ }\textbf {\bibinfo {volume}
  {B545}},\ \bibinfo {pages} {434} (\bibinfo {year} {1999})},\ \Eprint
  {http://arxiv.org/abs/hep-th/9808097} {arXiv:hep-th/9808097 [hep-th]}
  \BibitemShut {NoStop}%
%%CITATION = HEP-TH/9808097;%%
\bibitem [{\citenamefont {Onozawa}\ \emph {et~al.}(1996)\citenamefont
  {Onozawa}, \citenamefont {Mishima}, \citenamefont {Okamura},\ and\
  \citenamefont {Ishihara}}]{Onozawa:1995vu}%
  \BibitemOpen
  \bibfield  {author} {\bibinfo {author} {\bibfnamefont {H.}~\bibnamefont
  {Onozawa}}, \bibinfo {author} {\bibfnamefont {T.}~\bibnamefont {Mishima}},
  \bibinfo {author} {\bibfnamefont {T.}~\bibnamefont {Okamura}}, \ and\
  \bibinfo {author} {\bibfnamefont {H.}~\bibnamefont {Ishihara}},\ }\href
  {\doibase 10.1103/PhysRevD.53.7033} {\bibfield  {journal} {\bibinfo
  {journal} {Phys.Rev.}\ }\textbf {\bibinfo {volume} {D53}},\ \bibinfo {pages}
  {7033} (\bibinfo {year} {1996})},\ \Eprint
  {http://arxiv.org/abs/gr-qc/9603021} {arXiv:gr-qc/9603021 [gr-qc]}
  \BibitemShut {NoStop}%
%%CITATION = GR-QC/9603021;%%
\bibitem [{\citenamefont {Kallosh}\ \emph {et~al.}(1998)\citenamefont
  {Kallosh}, \citenamefont {Rahmfeld},\ and\ \citenamefont
  {Wong}}]{Kallosh:1997ug}%
  \BibitemOpen
  \bibfield  {author} {\bibinfo {author} {\bibfnamefont {R.}~\bibnamefont
  {Kallosh}}, \bibinfo {author} {\bibfnamefont {J.}~\bibnamefont {Rahmfeld}}, \
  and\ \bibinfo {author} {\bibfnamefont {W.~K.}\ \bibnamefont {Wong}},\ }\href
  {\doibase 10.1103/PhysRevD.57.1063} {\bibfield  {journal} {\bibinfo
  {journal} {Phys.Rev.}\ }\textbf {\bibinfo {volume} {D57}},\ \bibinfo {pages}
  {1063} (\bibinfo {year} {1998})},\ \Eprint
  {http://arxiv.org/abs/hep-th/9706048} {arXiv:hep-th/9706048 [hep-th]}
  \BibitemShut {NoStop}%
%%CITATION = HEP-TH/9706048;%%
\bibitem [{\citenamefont {Moncrief}(1975)}]{Moncrief:1975sb}%
  \BibitemOpen
  \bibfield  {author} {\bibinfo {author} {\bibfnamefont {V.}~\bibnamefont
  {Moncrief}},\ }\href {\doibase 10.1103/PhysRevD.12.1526} {\bibfield
  {journal} {\bibinfo  {journal} {Phys.Rev.}\ }\textbf {\bibinfo {volume}
  {D12}},\ \bibinfo {pages} {1526} (\bibinfo {year} {1975})}\BibitemShut
  {NoStop}%
%%CITATION = PHRVA,D12,1526;%%
\end{thebibliography}%
%%%
\end{document}